\documentclass[12pt]{article}

\usepackage[dvipdfmx]{graphicx}
\usepackage{amsfonts}
\usepackage[mathscr]{eucal}
\usepackage{ascmac}
\usepackage{dcolumn}
\usepackage{bm}
\usepackage[colorlinks=true,linkcolor=blue,citecolor=blue]{hyperref}
\usepackage{hyperref}
\usepackage{color}

\begin{document}


\newcommand{\gtrsim}{ \mathop{}_{\textstyle \sim}^{\textstyle >} }
\newcommand{\lesssim}{ \mathop{}_{\textstyle \sim}^{\textstyle <} }
\newcommand{\vev}[1]{ \left\langle {#1} \right\rangle }
\newcommand{\bra}[1]{ \langle {#1} | }
\newcommand{\ket}[1]{ | {#1} \rangle }
\newcommand{\EV}{ \ {\rm eV} }
\newcommand{\KEV}{ \ {\rm keV} }
\newcommand{\MEV}{\  {\rm MeV} }
\newcommand{\GEV}{\  {\rm GeV} }
\newcommand{\TEV}{\  {\rm TeV} }
\newcommand{\1}{\mbox{1}\hspace{-0.25em}\mbox{l}}
\newcommand{\Red}[1]{{\color{red} {#1}}}
\newcommand{\text}[1]{{\rm {#1}}}

\def\a{\alpha}
\def\b{\beta}
\def\c{\varepsilon}
\def\d{\delta}
\def\e{\epsilon}
\def\f{\phi}
\def\g{\gamma}
\def\h{\theta}
\def\k{\kappa}
\def\l{\lambda}
\def\m{\mu}
\def\n{\nu}
\def\p{\psi}
\def\q{\partial}
\def\r{\rho}
\def\s{\sigma}
\def\t{\tau}
\def\u{\upsilon}
\def\v{\varphi}
\def\w{\omega}
\def\x{\xi}
\def\y{\eta}
\def\z{\zeta}
\def\D{\Delta}
\def\G{\Gamma}
\def\H{\Theta}
\def\L{\Lambda}
\def\F{\Phi}
\def\P{\Psi}
\def\S{\Sigma}

\def\o{\over}
\def\beq{\begin{eqnarray}}
\def\eeq{\end{eqnarray}}
\newcommand{\gsim}{ \mathop{}_{\textstyle \sim}^{\textstyle >} }
\newcommand{\lsim}{ \mathop{}_{\textstyle \sim}^{\textstyle <} }
\def\diag{\mathop{\rm diag}\nolimits}
\def\Spin{\mathop{\rm Spin}}
\def\SO{\mathop{\rm SO}}
\def\O{\mathop{\rm O}}
\def\SU{\mathop{\rm SU}}
\def\U{\mathop{\rm U}}
\def\Sp{\mathop{\rm Sp}}
\def\SL{\mathop{\rm SL}}
\def\tr{\mathop{\rm tr}}

\def\IJMP{Int.~J.~Mod.~Phys. }
\def\MPL{Mod.~Phys.~Lett. }
\def\NP{Nucl.~Phys. }
\def\PL{Phys.~Lett. }
\def\PR{Phys.~Rev. }
\def\PRL{Phys.~Rev.~Lett. }
\def\PTP{Prog.~Theor.~Phys. }
\def\ZP{Z.~Phys. }

\def\dd{\mathrm{d}}
\def\ff{\mathrm{f}}
\def\BH{{\rm BH}}
\def\inf{{\rm inf}}
\def\ev{{\rm evap}}
\def\eq{{\rm eq}}
\def\SM{{\rm sm}}
\def\Mpl{M_{\rm Pl}}
\def\GeV{ \ {\rm GeV}}

\def\mDM{m_{\rm DM}}
\def\mphi{m_{\text{I}}}
\def\TeV{\ {\rm TeV}}
\def\MeV{\ {\rm MeV}}
\def\Gphi{\Gamma_\phi}
\def\TR{T_{\rm RH}}
\def\Br{{\rm Br}}
\def\DM{{\rm DM}}
\def\Eth{E_{\rm th}}
\newcommand{\lmk}{\left(}  
\newcommand{\rmk}{\right)}
\newcommand{\lkk}{\left[}  
\newcommand{\rkk}{\right]}
\newcommand{\lhk}{\left \{ }  
\newcommand{\rhk}{\right \} }
\newcommand{\del}{\partial}  
\newcommand{\la}{\left\langle} 
\newcommand{\ra}{\right\rangle}

\newcommand{\qel}{\hat{q}_{\rm el}}
\newcommand{\ksplit}{k_{\text{split}}}
\def\GDM{\Gamma_{\text{DM}}}
\newcommand{\half}{\frac{1}{2}}
\def\Gsplit{\Gamma_{\text{split}}}

\def\mg{m_{3/2}}
\newcommand{\abs}[1]{\left\vert {#1} \right\vert}
\def\Im{{\rm Im}}
\def\bea{\begin{array}}
\def\eea{\end{array}}
\def\Mpl{M_{\text{Pl}}}
\def\M{M_{\text{Pl}}}
\def\mN{m_{\text{NLSP}}}
\def\Td{T_{\text{decay}}}
\def\mphi{m_{\phi}}
\def\tanb{\text{tan}\beta}
\def\signmu{\text{sign}[\mu]}
\def\fb{\text{ fb}}
\def\ij{_{ij}}
\def\k{\lmk {\bf k} \rmk}
\def\tk{\lmk \tau, {\bf k} \rmk}
\def\xk{\lmk x, {\bf k} \rmk}
\def\TT{T_{ij}^{\rm TT}}
\def\Hz{\ {\rm Hz}}
\def\for{\quad \text{for }}
\def\Min{\text{Min}}
\def\Max{\text{Max}}
\def\Kahler{K\"{a}hler }
\def\cphi{\varphi}
\def\cm{\ {\rm cm}}
\def\km{\ {\rm km}}
\def\Mpc{\ {\rm Mpc}}


\begin{titlepage}

\baselineskip 8mm

\begin{flushright}
IPMU 15-0195; DESY 15-215
\end{flushright}

\begin{center}

\vskip 1.2cm

{\Large\bf
Affleck-Dine baryogenesis just after inflation 
}

\vskip 1.8cm

{\large 
Masaki Yamada$^{a,b,c}$
}

\vskip 0.4cm

{\it$^a$
ICRR, 
The University of Tokyo,
Kashiwa, Chiba 277-8582, Japan}\\
{\it$^b$Kavli IPMU (WPI), UTIAS, The University of Tokyo, 
Kashiwa, Chiba 277-8583, Japan}\\
{\it$^c$Deutsches Elektronen-Synchrotron DESY, 
22607 Hamburg, Germany}

\date{\today}
\vspace{2cm}

\begin{abstract}  
We propose a new scenario of Affleck-Dine baryogenesis 
where a flat direction in the MSSM generates $B-L$ asymmetry 
just after the end of inflation. 
The resulting amount of baryon asymmetry is independent of low-energy supersymmetric models 
but is dependent on inflation models. 
We consider the hybrid and chaotic inflation models 
and find that 
reheating temperature is required to be higher than that in the conventional scenario of 
Affleck-Dine baryogenesis. 
In particular, 
non-thermal gravitino-overproduction problem is naturally avoided 
in the hybrid inflation model. 
Our results imply that Affleck-Dine baryogenesis can be realized 
in a broader range of supersymmetry and inflation models than expected in the literature. 
\end{abstract}

\end{center}
\end{titlepage}

\baselineskip 6mm

\section{\label{introduction}Introduction}

The Big Bang theory is successful to explain 
the expansion of the Universe, 
the cosmic microwave background (CMB), 
and light element abundances. 
It requires the baryon-to-entropy ratio of order $10^{-10}$ 
as an initial condition at a temperature above $1 \MEV$. 
When we consider the earlier Universe, 
there is an era of exponential expansion, called inflation, 
which solves cosmological problems 
related to the initial conditions of the Universe, such as the horizon problem, 
flatness problem, and origin of large scale structure. 
However, 
baryon asymmetry is washed out by inflation, 
so that 
we need a mechanism to generate the observed amount of baryon asymmetry 
after inflation.

In supersymmetric (SUSY) theories, 
baryon asymmetry can be generated by Affleck-Dine baryogenesis (ADBG) 
using a $B-L$ charged flat direction called an AD field~\cite{AD, DRT}. 
The AD field is assumed to have a negative effective mass term, called a Hubble-induced mass term, 
due to a finite energy density of the Universe 
via supergravity effects, 
which implies that it obtains a large VEV 
during and after inflation. 
As the energy density of the Universe decreases, 
the effective mass decreases. 
Eventually, the effective mass becomes comparable to the soft mass of the AD field, 
and then the AD field starts to oscillate around the origin of its potential. 
At the same time, 
its phase direction is kicked by its A-term potential. 
Since the $B-L$ number density is proportional to the phase velocity of the AD field, 
the $B-L$ asymmetry is generated through this dynamics. 
Finally, the coherent oscillation of the AD field decays and dissipates into the thermal plasma 
and the $B-L$ asymmetry is converted to the desired baryon asymmetry 
through the sphaleron effects~\cite{Kuzmin:1985mm, Fukugita:1986hr}. 
There are many applications of ADBG (e.g., Refs.~\cite{Anisimov:2000wx, 
Fujii:2001zr, Mazumdar:2001nw, 
Baer:2009ms, Choi:2011rs, Kasuya:2014bxa, Yamada:2015rza}). 
It could solve the baryon-DM coincidence problem~\cite{EnMc, 
Fujii:2001xp, Roszkowski:2006kw, Kitano:2008tk, ShKu2, 
Kane:2011ih, Kamada:2012bk, Harigaya:2014tla, Kawasaki:2015cla} 
and the moduli problem~\cite{Felder:2007iz, Kawasaki:2007yy, Choi:2009qd, Furuuchi:2011wa, Higaki:2012ba, Garcia:2013bha, Hayakawa:2015fga}. 
The mechanism can also be used 
to generate asymmetry in dark sector~\cite{Bell:2011tn, Cheung:2011if, Fischler:2014jda}. 
Inflaton may play a role of the AD field in non-SUSY models~\cite{Hertzberg:2013mba, Hertzberg:2013jba}.

As mentioned above, the AD field obtains a 
Hubble-induced mass 
due to the finite energy density of the Universe during and after inflation 
(see Refs.~\cite{Kasuya:2006wf, Dutta:2010sg, Marsh:2011ud, Dutta:2012mw}
for recent works on Hubble-induced terms.) 
In the conventional scenario of ADBG, 
the sign of the Hubble-induced mass term is assumed to be negative 
during and after inflation. 
However, 
the sign of the Hubble-induced mass term can change after inflation 
because the source of the energy density of the Universe 
generically changes after inflation. 
In this paper, we investigate a new scenario 
that the AD field obtains a negative Hubble-induced mass term during inflation 
while it obtains a positive one after inflation.%
\footnote{
A similar scenario has been 
considered in the case of D-term inflation models in Refs.~\cite{McDonald:1999nc, Kawasaki:2015cla}, 
where the Hubble-induced mass is absent during D-term inflation 
and arises with a positive coefficient after inflation. 
In this paper, we focus on F-term hybrid and chaotic inflation models. 
}%
\footnote{
The opposite case, 
where the Hubble-induced mass term is positive during inflation and is negative after inflation, 
has been considered in Refs.~\cite{Kamada:2014qja, Kamada:2015iga}. 
Although $B-L$ asymmetry cannot be generated via the dynamics of the flat direction, 
topological defects form after inflation and emit gravitational waves. 
}
In this case, the AD field 
starts to oscillate around the origin of the potential due to the positive Hubble-induced mass term 
just after the end of inflation. 
At the same time, its phase direction 
is kicked by an A-term 
and 
$B-L$ asymmetry is generated. 
We calculate the produced amount of baryon asymmetry and show that 
it can be consistent with that observed.

The whole scenario is much simpler than the conventional scenario of ADBG. 
This is because the dynamics of the AD field is determined only by the Hubble-induced terms 
and the low-energy potential of the AD field does not affect the resulting $B-L$ asymmetry. 
This means that the scenario and our calculations in this paper 
can be applied to many SUSY models, 
including gravity-mediated and gauge-mediated SUSY breaking models. 
In particular, the scenario does not result in the formation of non-topological solitons called Q-balls 
even in gauge-mediated SUSY breaking models~\cite{Coleman, Qsusy, EnMc, KuSh, KK}. 
This is one of the advantages of our scenario 
because Q-balls are sometimes problematic due to their long lifetime. 
In addition, thermal effect on the dynamics of the AD field can be neglected 
in our scenario 
because it starts to oscillate before thermal plasma grows. 
This is the case even for so-called $L H_u$ flat direction. 
However, 
the resulting $B-L$ asymmetry depends on 
the energy scale of inflation 
because the dynamics of the AD field is determined by Hubble-induced terms. 
In particular, 
the A-term depends on inflation models, 
so that 
we need to calculate $B-L$ asymmetry for each inflation model. 
Since the resulting $B-L$ asymmetry depends on parameters in inflaton sector, 
we could check the consistency of the scenario 
by observing predictions of inflation models, such as the spectral index and tensor-to-scalar ratio.

This paper is organised as follows.
In the next section, we briefly review the conventional scenarios of ADBG. 
Then we consider our scenario of ADBG 
in the case that the AD field obtains 
a positive Hubble-induced mass term after inflation. 
We first overview the scenario in Sec.~\ref{ADBG just after inf}. 
Then we 
apply it to a hybrid inflation model in Sec.~\ref{hybrid} 
and a chaotic inflation model in Sec.~\ref{chaotic}. 
Finally, we conclude in Sec.~\ref{conclusion}.

\section{\label{ADBG}Conventional scenario of ADBG}

In this section, we review the conventional scenario of ADBG 
to clarify the difference from our scenario explained in the subsequent sections.

\subsection{Preliminary}

In SUSY theories, 
there are SUSY partners of quarks and leptons, called squarks and sleptons, 
which are complex scalar fields carrying $B-L$ charges. 
Let us consider one of them and denote it as $\phi$. 
When we write its $B-L$ charge as $q$, 
the number density of $B-L$ asymmetry associated with $\phi$ 
is written as 
\beq
 n_{B-L} = i q \lmk \dot{\phi}^* \phi - \phi^* \dot{\phi} \rmk = 2 q \Im \lkk \phi^* \dot{\phi} \rkk. 
 \label{n_B-L}
\eeq
This implies that we can obtain a large amount of $B-L$ asymmetry 
when the field $\phi$ rotates in the complex plane with a large amplitude. 
Thus 
we focus on a $B-L$ charged scalar field that has a very flat potential. 
In SUSY theories, 
there are two types of potentials for scalar fields: D-term and F-term potentials. 
Although gauged scalar fields have D-term potentials, 
it is known that D-terms are cancelled for gauge-singlet combinations of scalar fields. 
For example, 
when the field $\phi$ consists of the following combination, 
D-term potentials are cancelled: 
\beq
 (u^c)^R_i = \frac{1}{\sqrt{3}} \phi, \quad 
 (d^c)^G_j = \frac{1}{\sqrt{3}} \phi, \quad
 (d^c)^B_k = \frac{1}{\sqrt{3}} \phi, 
\eeq
where the upper indices represent color 
and the lower ones represent flavours ($j \ne k$). 
The fields $u^c$ and $d^c$ are $u$-type and $d$-type right-handed squarks, respectively. 
This D-flat direction is sometimes called $u^c d^c d^c$ flat direction. 
The following combination is 
another famous example of flat directions 
called $L H_u$ flat direction~\cite{Murayama:1993em}: 
\beq
 L_i = \frac{1}{\sqrt{2}} 
 \lmk 
 \bea{ll}
 0
 \\
 \phi
 \eea
 \rmk, 
 \quad 
 H_u = \frac{1}{\sqrt{2}} 
  \lmk 
 \bea{ll}
 \phi
 \\
  0
 \eea
 \rmk, 
\eeq
where $L$ and $H_u$ are left-handed slepton and up-type Higgs, respectively. 
F-term potentials are determined by superpotential $W$ as 
\beq
V_F (\phi) = \abs{\frac{\del W}{\del \phi} }^2. 
\eeq
In the minimal SUSY Standard Model (MSSM), the superpotential is given by 
\beq
 W^{(\rm MSSM)} = y_u Q H_u u^c - y_d Q H_d d^c - y_e L H_d e^c + \mu H_u H_d, 
 \label{W_MSSM}
\eeq
within the renormalizable level, 
where we omit flavour indices. 
Here we implicitly assume R-parity conservation to avoid disastrous proton decay. 
Fortunately, many D-flat directions, including $u^c d^c d^c$ flat direction, 
have no F-term potential within the renormalizable level. 
The D- and F-flat directions with nonzero $B-L$ charge is listed in Table.~\ref{table}~\cite{Gherghetta:1995dv}.%
\footnote{
Although $L H_u$ flat direction has a potential coming from the Higgs $\mu$-term, 
it is assumed that $\mu$ is of order the soft mass scale and absorb it to the meaning of $m_\phi$ [see 
Eq.~(\ref{A-term})]. 
}
It is expected that the dynamics of such a flat direction can generate 
a large amount of $B-L$ asymmetry.

\begin{table}
\caption{\label{table}
Flat directions in the MSSM and $B-L$ charges~\cite{Gherghetta:1995dv}. 
}
\begin{center}
\begin{tabular}{ll}
 flat directions & $B-L$ 
\\
 \hline \hline
$ L H_u $  & -1  \\  \hline
$ u^c d^c d^c $  & -1  \\  \hline
$ LLe^c $  & -1   \\ \hline
$ Qd^c L $  & -1   \\  \hline
$d^cd^cd^cLL  $ & -3   \\  \hline
$ u^cu^cu^ce^ce^c $  & 1  \\  \hline
$ Qu^cQu^ce^c $  & 1  \\  \hline
$ QQQQu^c $  & 1  \\  \hline
$ (QQQ)_4 LLLe^c $  & -1  \\  \hline
$ u^cu^cd^c Qd^cQd^c $  & -1 \\   \hline
\end{tabular}
\end{center}
\end{table}

In low energy, 
the AD field obtains soft terms coming from the low-energy SUSY breaking effect. 
In this section, 
we consider gravity-mediated SUSY breaking models for simplicity. 
Note that the conventional scenario of ADBG 
depends on mediation models,%
\footnote{
When we consider a SUSY model with a gauge mediated SUSY breaking effect, 
the soft mass of the AD field is suppressed for a VEV larger than the messenger scale~\cite{deGouvea:1997tn}. 
In this case, we have to take into account the formation of non-topological solitons called Q-balls~\cite{
Coleman, Qsusy, EnMc, KuSh, KK}. 
The baryon number should be released from Q-balls 
to explain the observed amount of baryon asymmetry and the scenario is completely different from the one 
explained in this section~\cite{Fujii:2001xp, ShKu2, Kamada:2012bk, Harigaya:2014tla}. 
} 
but our scenario does not as explained in the subsequent sections. 
We write soft terms of the AD field as 
\beq
 V_{\rm soft} &=& m_\phi^2 \abs{\phi}^2 
 + a m_{3/2} W^{(\rm AD)} + {\rm c.c.} 
 \label{A-term} 
\eeq
where $m_\phi$ ($\simeq m_{3/2}$) is the soft mass of the AD field, 
$m_{3/2}$ is gravitino mass, 
and $a$ ($= \mathcal{O}(1)$) is a constant. 
We can assume $a = a^*$ without loss of generality. 
The higher-dimentional superpotential of the AD field $W^{(\rm AD)}$ is determined below.

During and after inflation, 
the AD field obtains effective potentials 
from the energy density of inflaton $I$ via supergravity effects. 
In supergravity, the potential of scalar fields is determined by 
\beq
 V_{\rm SUGRA} = e^{K/\Mpl^2} \lkk \lmk D_i W \rmk K^{i \bar{j}} \lmk D_j W \rmk^* - \frac{3}{\Mpl^2} \abs{W}^2 \rkk, 
 \label{SUGRA potential}
\eeq
where $K$ is a \Kahler potential 
and 
$D_i W \equiv W_i + K_i W / \Mpl^2$. 
The subscripts represent the derivatives with respect to corresponding fields, 
e.g., $W_i = \del W / \del \phi$ for $i = \phi$, 
and $K^{i \bar{j}}$ is defined by the inverse of $K_{i \bar{j}}$. 
We introduce an inflaton $I$ 
with a \Kahler potential of 
\beq
 K = \abs{\phi}^2 + \abs{I}^2 + \frac{c}{\Mpl^2} \abs{\phi}^2 \abs{I}^2, 
\eeq
where 
$c$ is an $O(1)$ constant. 
We assume that the F-term potential of $I$ drives inflation and satisfies $\abs{W_I}^2 \simeq 3 H_{\rm inf}^2 \Mpl^2$, 
where $H_{\rm inf}$ is the Hubble parameter during inflation. 
The supergravity potential of Eq.~(\ref{SUGRA potential}) 
includes 
the following interaction: 
\beq
 V &\supset& 
 \exp \lmk \frac{K}{\Mpl^2} \rmk 
 W_I ( K^{I \bar{I}})^{-1} W_I^* \\
 &\simeq& 
 \abs{F_I}^2 \lmk 1 + (1 - c) \abs{\phi}^2 \rmk, 
 \label{H-mass during inf}
\eeq
where we assume $\la \phi \ra, \la I \ra \ll \Mpl$ 
and neglect irrelevant higher-dimensional terms. 
Thus the AD field $\phi$ 
obtains an effective mass term of order the Hubble parameter during inflation: 
\beq
 V &\supset& c_H H_{\rm inf}^2 \abs{\phi}^2 \\
 c_H &=& - 3 (c - 1). 
\eeq
This is called a Hubble-induced mass term.%
\footnote{
There is sometimes a Hubble-induced A-term during inflation, 
but it is not the case in general (see Ref.~\cite{Kasuya:2008xp}). 
}

After inflation ends, 
the inflaton starts to oscillate around the potential minimum and 
its oscillation energy dominates the Universe. 
During this inflaton-oscillation dominated era, 
the Hubble-induced mass comes also from 
higher-dimensional kinetic interactions, which are determined by the \Kahler potential 
as 
\beq
 \mathcal{L}_{\rm kin} = K_{i \bar{j}} \del_\mu \cphi^i \del^\mu \cphi^{* j}, 
 \label{kinetic term}
\eeq
where $\cphi_i$ generically represents the fields of $\phi$ and $I$. 
There is a kinetic interaction of 
\beq
 \mathcal{L}_{\rm kin} \supset K_{I \bar{I}} \abs{\dot{I}}^2 
 \supset \frac{c}{\Mpl^2} \abs{\dot{I}}^2 \abs{\phi}^2. 
\eeq
A typical time scale of the dynamics of the AD field is at most of order the Hubble parameter 
as shown below. 
That of inflaton is the curvature of its potential, which is larger than the Hubble parameter 
during inflaton-oscillation dominated era. 
Thus we can 
take a time-average over the inflaton-oscillation time scale 
to investigate the dynamics of the AD field. 
Assuming that the inflaton oscillates in a quadratic potential after inflation, 
we obtain an effective Hubble-induced mass for $\phi$ after inflation: 
\beq
 V_H = c_H H^2 (t) \abs{\phi}^2 
 \\
 c_H = - 3 \lmk c - \frac{1}{2} \rmk, 
 \label{H-mass after inf}
\eeq
where we use the Virial theorem 
and 
include the contribution from the F-term potential.%
\footnote{
Inflation may be driven by a D-term potential of inflaton. 
In this case, 
the Hubble-induced mass is absent during inflation 
but 
the AD field stays at a nonzero VEV due to the Hubble-friction effect~\cite{Kolda:1998kc, Enqvist:1998pf, Kawasaki:2001in}. 
The inflaton obtains nonzero F-term 
after inflation ends, 
so that 
the AD field obtains a Hubble-induced mass during the inflaton oscillation dominated era. 
Thus the scenario of ADBG 
and resulting $B-L$ asymmetry are the same with the ones in F-term inflation. 
}

In the conventional scenario, 
$c_H$ is assumed to be negative during and after inflation. 
This means that the AD field has a large tachyonic mass 
and obtains a large VEV during the time of $H(t) \gtrsim m_\phi$. 
Since 
the AD field has a large VEV, 
we have to take into account non-renormalizable terms 
to investigate its dynamics. 
Although the superpotential of the AD field is absent within the renormalizable level, 
it may have a higher-dimensional superpotential such as 
\beq
 W^{(\rm AD)} = \lambda \frac{\phi^n}{n \Mpl^{n-3}}, 
 \label{W_AD}
\eeq
where $n$ ($ \ge 4$) is an integer depending on flat directions 
and $\Mpl$ ($\simeq 2.4 \times 10^{18} \GEV$) is the reduced Planck scale. 
For example, since the neutrinos have nonzero masses (denoted as $m_{\nu_i}$), 
we introduce a superpotential of 
\beq
 W^{(L H_u)} &=& \frac{m_{\nu_i}}{2 \la H_u \ra^2} \lmk L_i H_u \rmk^2, \\ 
 &\equiv& \frac{\lambda}{4 \Mpl} \phi^4
 ~~~~~~\text{for}~~~~ \frac{\phi^2}{2} = L H_u, 
\eeq
where $\la H_u \ra = \sin \beta \times 174 \GeV$ and $\tan \beta \equiv \la H_u \ra / \la H_d \ra$. 
Thus 
$L H_u$ flat direction corresponds to the case of $n=4$ in Eq.~(\ref{W_AD}). 
We can also write a superpotential of $(u^c d^c d^c)^2$, 
so that $n=6$ for the $u^c d^c d^c$ flat direction. 
The superpotential leads to 
a F-term potential of $\phi$ as 
\beq
 V_F (\phi) = \lambda^2 \frac{\abs{\phi}^{2n-2}}{\Mpl^{2n-6}}, 
 \label{V_F}
\eeq
where we neglect irrelevant higher-dimensional terms in the supergravity potential.

\subsection{Case without thermal effects}

Let us explain the dynamics of the AD field 
and calculate $B-L$ asymmetry. 
In this section, we neglect thermal log potential, which is 
explained and introduced in the next subsection.

As explained in the previous subsection, 
the potential of the AD field 
is given by 
\beq
 V (\phi) 
 &=& V_{\rm soft} + V_H + V_F 
 \\ 
 &=& m_\phi^2 \abs{\phi}^2 
 + \lmk a m_{3/2} \lambda \frac{\phi^n}{n \Mpl^{n-3}} + {\rm c.c.}  \rmk
 + c_H H^2(t) \abs{\phi}^2 
 + \lambda^2 \frac{\abs{\phi}^{2n-2}}{\Mpl^{2n-6}}, 
\eeq
during the inflaton-oscillation dominated era. 
When we decompose the AD field as $\phi = \cphi e^{i \theta} / \sqrt{2}$, 
the equations of motion 
are written as 
\beq
 \ddot{\cphi} + 3 H \dot{\cphi} - \dot{\theta}^2 \cphi + \frac{\del V(\cphi)}{\del \cphi} &=& 0
\\
 \ddot{\theta} + 3 H \dot{\theta} + 2 \frac{\dot{\cphi}}{\cphi} \dot{\theta} 
 + 
 \frac{\del V}{\del \theta} 
 &=& 0, 
 \label{EOM for phase direction}
\eeq 
where $H = 2/3 t$ during the inflaton-oscillation dominated era. 
Note that the phase direction has a Hubble-friction term ($3 H \dot{\theta}$).

The coefficient $c_H$ is assumed to be negative in the conventional scenario of ADBG. 
In this case, 
the AD field has a tachyonic mass, 
so that it obtains a large VEV. 
The VEV of the AD field at the potential minimum is given by 
\beq
 \la \abs{\phi} \ra \simeq 
 \lmk \frac{\abs{c_H} H^2(t) \Mpl^{2n-6}}{\lambda^2 (n-1)} \rmk^{1/(2n-4)}. 
 \label{VEV}
\eeq
for $H(t) \gtrsim m_\phi$. 
The AD field follows this potential minimum.

\begin{figure}[t]
\centering 
\begin{tabular}{l l}
\includegraphics[width=.45\textwidth, bb=0 0 450 305
]{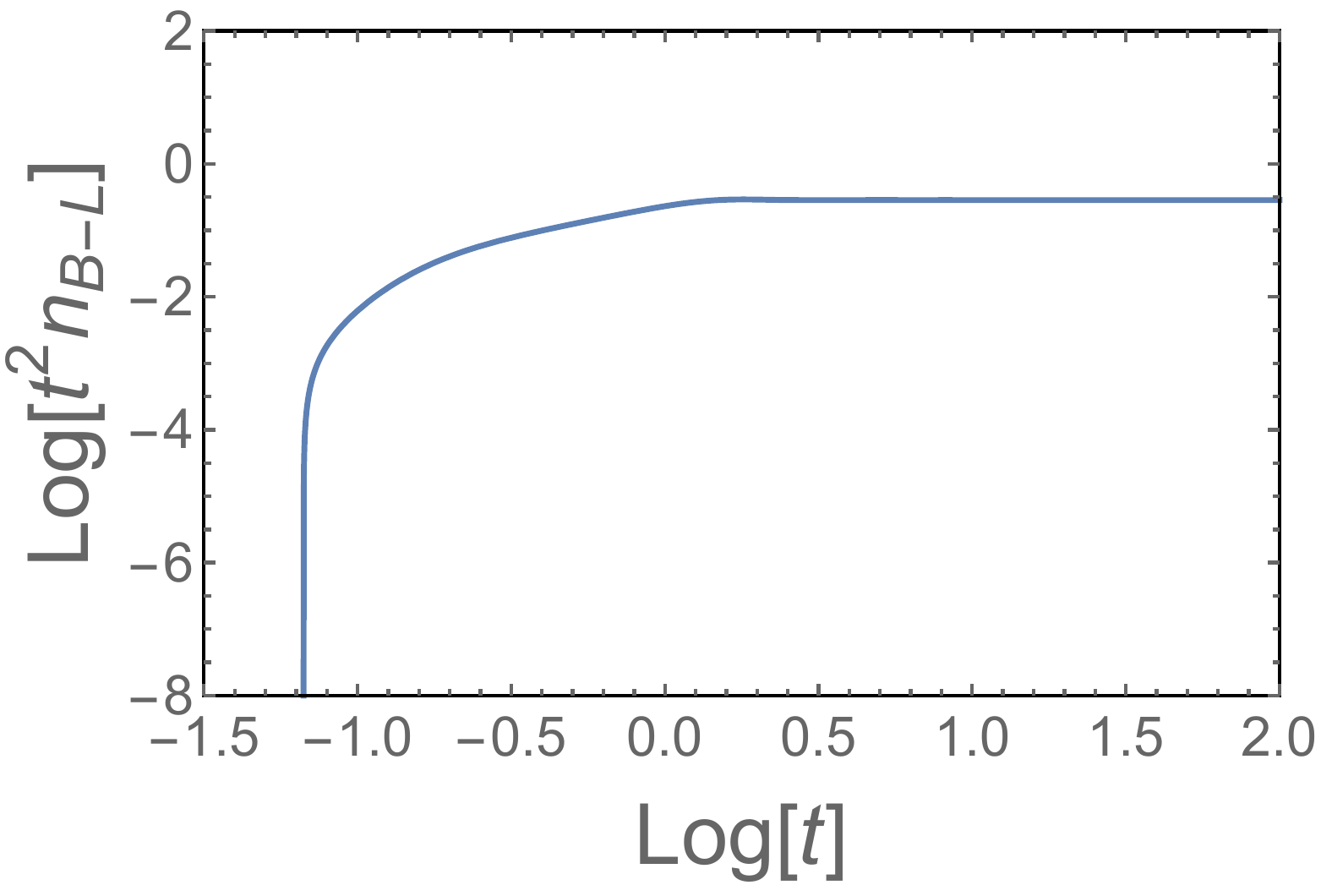}
\quad 
\includegraphics[width=.45\textwidth, bb=0 0 450 191
]{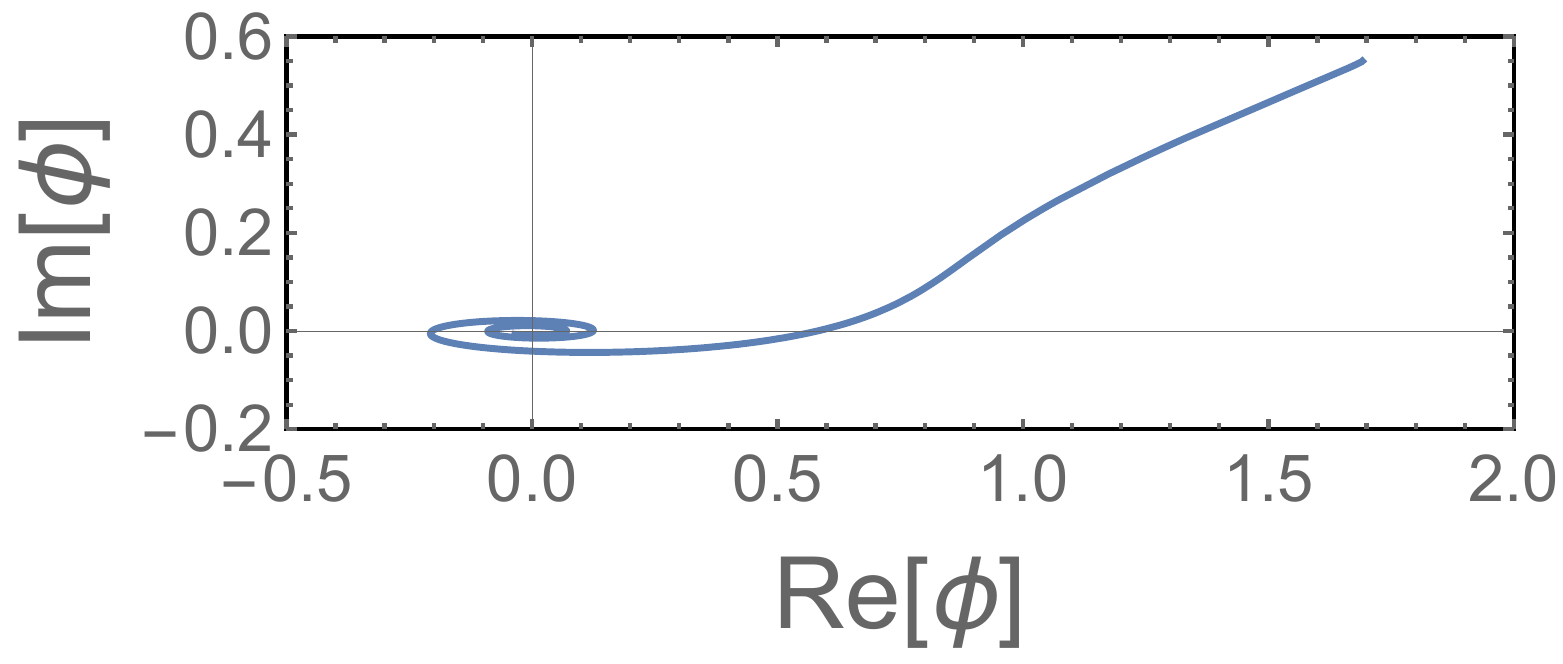} 
\end{tabular}
\caption{\small
Evolution of $B-L$ number density in a comoving volume (left panel) and the AD field (right panel) 
in the conventional scenario of ADBG. 
We set $n=6$, $c_H = -1$, $a m_{3/2} / m_\phi = -1$, 
and $\theta_0 = \pi / 10$. 
The dimensionfull quantities are rescaled such as $t \to t/m_\phi$ and $\phi \to \phi / \la \abs{\phi} \ra_{t=H_{\rm osc}^{-1}}$. 
}
  \label{fig1}
\end{figure}

The phase of the flat direction stays at a certain phase 
due to the Hubble friction term. 
We denote the initial phase of the AD field as $\theta_0$, 
which is expected to be of order unity. 
When the Hubble parameter decreases to $m_\phi$, 
the potential of the AD field is dominated by the soft mass term and 
it starts to oscillate around the origin of the potential. 
Here we denote the Hubble parameter at the time of beginning of oscillation as $H_{\rm osc}$: 
\beq
 H_{\rm osc} \simeq \frac{m_\phi}{\sqrt{\abs{c_H}}}. 
 \label{H_osc 1}
\eeq
The VEV of the AD field at that time is given by 
\beq
 \phi_{\rm osc} \simeq 
  \lmk \frac{\abs{c_H} H^2_{\rm osc} \Mpl^{2n-6}}{\lambda^2 (n-1)} \rmk^{1/(2n-4)}. 
  \label{VEV2}
\eeq
At the same time, 
its phase direction is kicked by the A-term, 
so that 
it starts to rotate in the phase space. 
This is the dynamics that generates the $B-L$ asymmetry [see Eq.~(\ref{n_B-L})]. 
The evolution of equation for the $B-L$ number density is written as 
\beq
 \dot{n}_{B-L} +3 H n_{B-L} 
 = - q \cphi^2 \lmk \frac{\del V}{\del \theta} \rmk, 
\eeq
where $q$ denotes the $B-L$ charge of the AD field. 
We semi-analytically and numerically solve this equation and obtain 
\beq
  a^3 n_{B-L} (t) 
 &=& 
 - \int \dd t q a^3(t) \cphi^2 \frac{\del V}{\del \theta} \\
 &\equiv& 
 \epsilon q H_{\rm osc} \phi_{\rm osc}^2 a^3 (t_{\rm osc}) 
 \\
 \epsilon 
 &\simeq& 
  (2-4) \times \frac{a}{\sqrt{n-1}(1+(n-4)/(n-2))} \frac{m_{3/2}}{m_\phi} \sin \lmk n \theta_0 \rmk 
 ~~~~ \text{ for }~~ \epsilon \lesssim 1, 
 \label{result in conventional scenario 1}
\eeq
where 
we assume $c_H = -1$ in the last line. 
We define the ellipticity parameter $\epsilon$ ($\le 1$) which represents the efficiency of baryogenesis. 
Since the $B-L$ number density has to be smaller than that of the total AD field times $B-L$ charge $q$, 
$\epsilon$ is at most unity. 
We have numerically solved the equation of motion for $\phi$ 
and have obtained the numerical factor of $(2-4)$ in Eq.~(\ref{result in conventional scenario 1}) 
for $c_H = -1$ and $\epsilon \lesssim 1$. 
One of the numerical results is shown in Fig.~\ref{fig1}, 
where we set $n=6$, $c_H = -1$, $a m_{3/2} / m_\phi = -1$, 
and $\theta_0 = \pi / 10$. 
One can see that the phase direction is kicked and 
the $B-L$ asymmetry is generated at $t \sim m_\phi^{-1} \simeq H_{\rm osc}^{-1}$. 
The amplitude of the flat direction decreases as time evolves 
due to the Hubble expansion 
and the $B-L$ breaking effect (i.e., the A-term) becomes irrelevant soon after the oscillation. 
Thus, the generated $B-L$ asymmetry within a comoving volume 
is conserved soon after the AD field starts to oscillate 
as one can see in Fig.~\ref{fig1}.

Then, the oscillating AD field decays and dissipates into radiation~\cite{Mukaida:2012qn} 
and the sphaleron effect relates 
the $B-L$ asymmetry to the baryon asymmetry~\cite{Kuzmin:1985mm, Fukugita:1986hr}.%
\footnote{
For simplicity, 
in this section 
we assume that Q-balls do not form after ADBG. 
Note that in our scenario explained in the subsequent sections, 
Q-balls do not form. 
}
Since the sphaleron process is in thermal equilibrium, 
the resulting baryon asymmetry is related to the $B-L$ asymmetry such as~\cite{Harvey:1990qw}
\beq
 n_b \simeq \frac{8}{23} 
 n_{B-L}. 
\eeq
We can calculate the resulting baryon-to-entropy ratio $Y_b$ such as 
\beq
 Y_b  
 &\equiv& 
 \frac{n_b}{s} 
 \simeq 
 \left. \frac{8}{23} \frac{n_{B-L}}{s} \right\vert_{\rm RH} \\
 &\simeq& 
 \left. \frac{8}{23} \frac{3 T_{\rm RH} n_{B-L}}{4 \rho_{\rm inf}} \right\vert_{\rm osc} \\
 &\simeq& 
 \frac{8}{23} \frac{\epsilon q T_{\rm RH}}{4 H_{\rm osc}} \lmk \frac{\phi_{\rm osc}}{\Mpl} \rmk^2 \\
 \label{Y_b 1}
 &\simeq& 
 1.2 \times 10^{-10} 
 \epsilon q \lambda^{-1/2} 
 \lmk \frac{T_{\rm RH}}{100 \GEV} \rmk 
 \lmk \frac{m_\phi}{1 \TEV} \rmk^{-1/2}~~~~\text{for}~~n=6, 
 \label{Y_b conventional}
\eeq
where $\rho_{\rm inf}$ ($\simeq 3 H^2(t) \Mpl^2$) is the energy density of the inflaton 
and $T_{\rm RH}$ is reheating temperature. 
In the last line, we use Eq.~(\ref{VEV2}). 
The resulting baryon asymmetry can be consistent with the observed baryon asymmetry of 
$Y_b^{(\rm obs)} \simeq 8.7 \times 10^{-11}$~\cite{pdg}. 
Since we expect $\epsilon q \sim 1$, 
a relatively low reheating temperature is required to explain the observed amount of baryon asymmetry 
unless the parameter $\lambda$ is much larger than unity.

\subsection{Case with thermal effects: $L H_u$ flat direction}

In this section, we take into account thermal log potential. 
It is particularly important for the case of $n=4$, 
including the case of $L H_u$ flat direction.

After inflation ends and before reheating completes, 
inflaton gradually decays into radiation.
Since 
the energy density of radiation is given by 
$\rho_{\rm rad} \simeq (3/5) \rho_{\rm inf} \Gamma_I t$, 
there is a background plasma with a temperature of 
\beq
 T = \lmk \frac{36 H(t) \Gamma_I \Mpl^2}{g_* (T) \pi^2} \rmk^{1/4}, 
 \label{T during osc.}
\eeq
where 
$g_*$ is the effective number of relativistic degrees of freedom in the thermal plasma. 
The decay rate of inflaton $\Gamma_I$ is related with the reheating temperature as 
\beq
 T_{\rm RH} \simeq \lmk \frac{90}{g_* (T_{\rm RH}) \pi^2 } \rmk^{1/4} \sqrt{\Gamma_I \Mpl}. 
\eeq

Here we explain the origin of the thermal log potential, 
focusing on $L H_u$ flat direction. 
The free energy of the thermal plasma $F$ depends on QCD coupling $g_s$ 
in the next-to-leading order 
as 
\beq
 F = \frac{3}{8} (1 + N_f^{({\rm th})}) g_s^2 (T) T^4, 
\eeq
where $N_f^{({\rm th})}$ is the number of family in the thermal plasma. 
Here, the quark multiplets obtain effective masses via the Yukawa interactions 
when $L H_u$ flat direction has a large VEV [see Eq.~(\ref{W_MSSM})]. 
When its VEV is larger than the temperature of the plasma, 
the renormalization running of $g_s$ is affected 
and 
its value at the energy scale of $T$ 
depends on the VEV of $L H_u$ flat direction: 
$g_s(T)  = g_s (T, \phi)$. 
Therefore 
the free energy depends on $\phi$ and 
$L H_u$ flat direction acquires a potential depending on temperature. 
Since the renormalization running has a logarithmic dependence, 
it is written as~\cite{Anisimov:2000wx, Fujii:2001zr} 
\beq
 V_T (\phi) \simeq c_T \alpha_s^2 T^4 \log \lmk \frac{\abs{\phi}^2}{T^2} \rmk, 
\eeq
with $c_T = 45/ 32$ for $y \abs{\phi} \gg T$, 
where $\alpha_s \equiv g_s^2 / 4 \pi$ and $y$ generically stands for Yukawa couplings for quarks. 
This is sometimes called thermal log potential.

In the previous subsection, 
we neglect the thermal potential and 
the AD field starts to oscillate around the origin of the potential at $H(t) \simeq m_\phi / \sqrt{\abs{c_H}}$. 
When we take into account the thermal log potential, 
it starts to oscillate at the time of 
\beq
 H_{\rm osc} \simeq \Max \lkk \frac{m_\phi}{\sqrt{\abs{c_H}}}, \ \sqrt{\phi^{-1} V_T'} \rkk. 
 \label{H_osc 1-2}
\eeq
Using Eqs.~(\ref{VEV2}) and (\ref{T during osc.}), 
this can be rewritten as 
\beq
 H_{\rm osc} \simeq 
 \Max 
 \lkk 
 m_\phi 
 , \ 
 0.6 
 \alpha_s \sqrt{\lambda} T_{\rm RH} 
 \rkk, 
\label{H_osc 2}
\eeq
where we 
assume $\abs{c_H} = 1$ and $n=4$.

\begin{figure}[t]
\centering 
\begin{tabular}{l l}
\includegraphics[width=.45\textwidth, bb=0 0 450 305
]{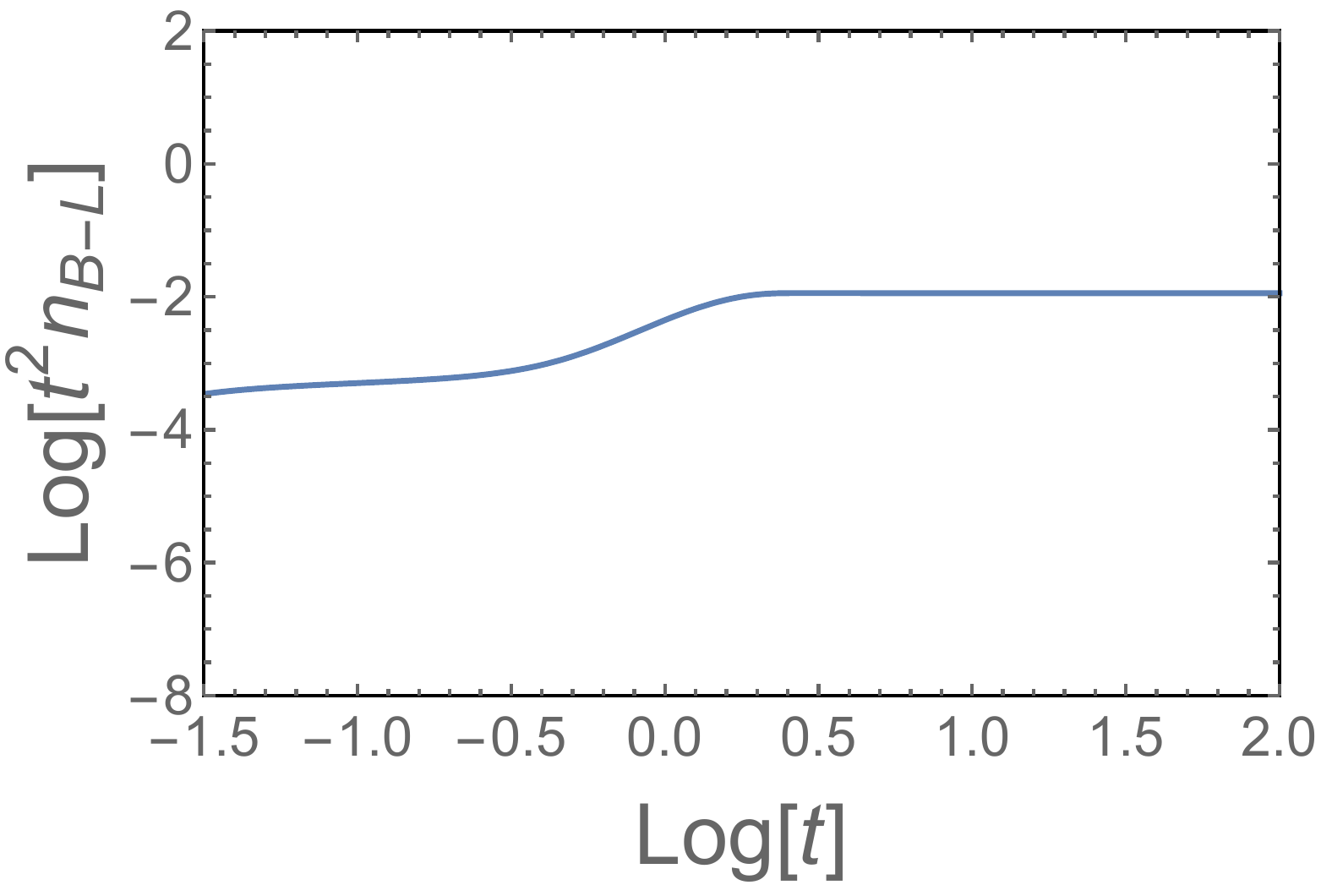} 
\quad 
\includegraphics[width=.45\textwidth, bb=0 0 450 191
]{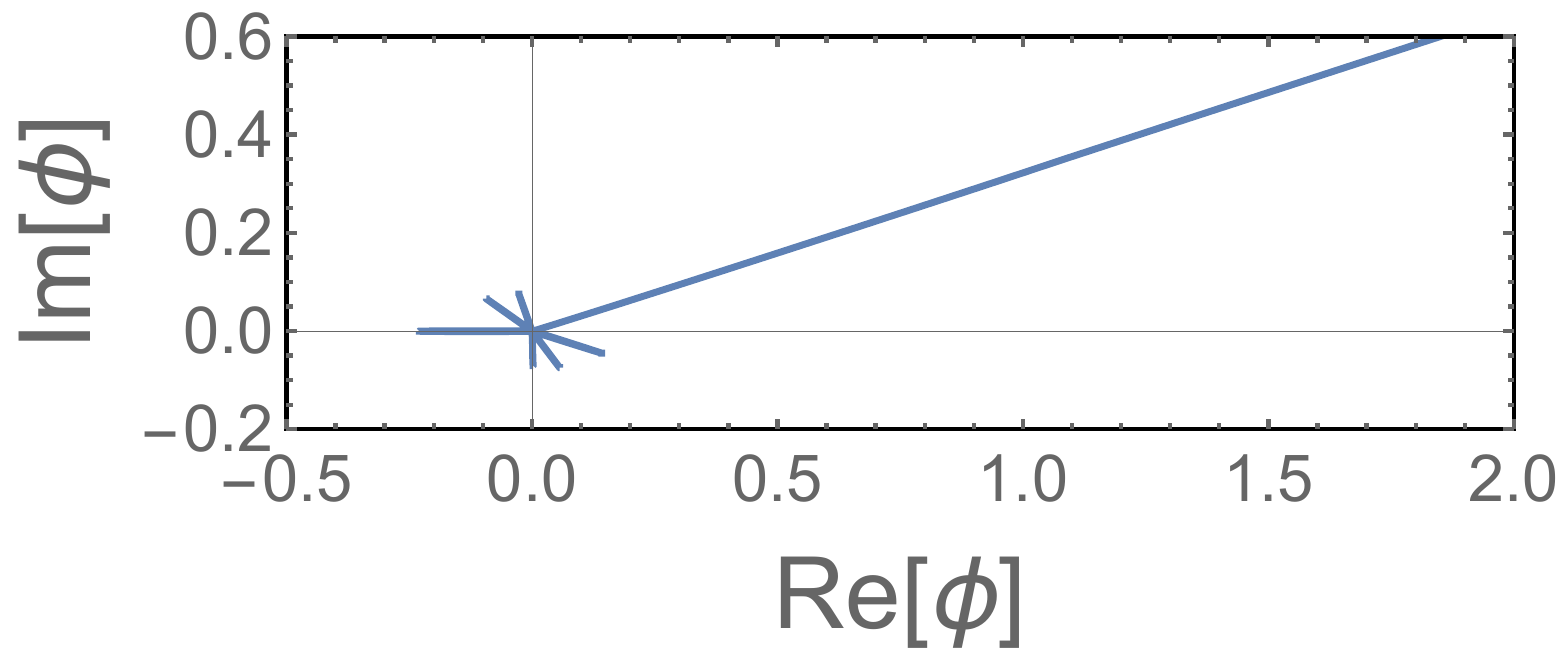} 
\end{tabular}
\caption{\small
Evolution of $B-L$ number density in a comoving volume (left panel) and the phase direction of the AD field (right panel) 
in the conventional scenario of ADBG. 
We set $c_H = -1$, $a_H m_{3/2} / H_{\rm osc} = -0.01$, 
and $\theta_0 = \pi / 10$. 
The dimensionfull parameters are rescaled as $t \to t/H_{\rm osc}$ and $\phi \to \phi / \la \abs{\phi} \ra_{t=H_{\rm osc}^{-1}}$. 
}
  \label{fig2}
\end{figure}

We numerically solve the equation of motion for $\phi$ 
and obtain the ellipticity parameter as 
\beq
 \epsilon &=& (0.4 - 3.5) \times a \sin \lmk n \theta_0 \rmk \frac{m_{3/2}}{H_{\rm osc}} 
 \label{result in conventional scenario 2}
 \\
 &\equiv& \tilde{\epsilon} \frac{m_{3/2}}{H_{\rm osc}}, 
\eeq
where we define $\tilde{\epsilon}$ that is expected to be of order unity. 
Here we assume 
$T_{\rm RH} \gtrsim m_\phi / (\alpha_s \sqrt{\lambda})$, 
which implies $H_{\rm osc} \simeq 0.6 \alpha_s \sqrt{\lambda} T_{\rm RH}$ [see Eq.~(\ref{H_osc 2})]. 
One of our results is shown in Fig.~\ref{fig2}, 
where we set $c_H = 1$, $a m_{3/2} / H_{\rm osc} = -0.01$, and $\theta_0 = \pi / 10$. 
The ellipticity parameter $\epsilon$ is much smaller than unity in this numerical calculation, 
so that the phase direction is kicked slightly. 
We are difficult to see that the AD field rotates in the phase space in the right panel of Fig.~\ref{fig2} 
though it actually does.

The baryon-to-entropy ratio is calculated as 
\beq
 Y_b &\simeq& 
 \frac{8}{23} \frac{q \tilde{\epsilon} m_{3/2}}{4 \alpha_s \lambda^{3/2} \Mpl} \\
 &\simeq& 
 3.7 \times 10^{-10} \tilde{\epsilon} \lmk \frac{\lambda}{10^{-4}} \rmk^{-3/2} 
 \lmk \frac{m_{3/2}}{1 \TEV} \rmk, 
 \label{Y_b 2}
\eeq
where we assume $T_{\rm RH} \gtrsim m_\phi / (\alpha_s \sqrt{\lambda})$, 
$\abs{c_H} = 1$, 
and $\alpha_s = 0.1$ 
and use $\epsilon = \tilde{\epsilon} m_{3/2} / H_{\rm osc}$. 
This result is independent of the reheating temperature~\cite{Fujii:2001zr}. 
The observed baryon asymmetry can be explained when the coupling $\lambda$ satisfies 
\beq
 \lambda \simeq 2.6 \times 10^{-4} \lmk \frac{m_{3/2}}{1 \TEV} \rmk^{2/3}, 
\eeq
where we assume $\tilde{\epsilon} = 1$. 
When we identify the AD field as $L H_u$ flat direction, 
this result implies that the lightest left-handed neutrino has a tiny mass of 
\beq
 m_{\nu} 
 &\simeq& 1.6 \times 10^{-9} \EV \lmk \frac{\lambda}{2.6 \times 10^{-4} } \rmk \\
 &\simeq& 1.6 \times 10^{-9} \EV \lmk \frac{m_{3/2}}{1 \TEV } \rmk^{2/3}. 
\eeq

\subsection{\label{isocurvature}Baryonic isocurvature constraint}

In many cases, 
the phase direction of the AD field is massless during inflation. 
This implies that the phase direction has a quantum fluctuations during inflation~\cite{Enqvist:1998pf, Kawasaki:2001in, Kasuya:2008xp, Harigaya:2014tla}:
\beq
 \abs{\delta \theta_0} \simeq \frac{\sqrt{2} H_{\rm inf} }{2 \pi  \abs{\phi}_{\rm inf}}. 
\eeq
Since the resulting baryon asymmetry is related to $\theta_0$ [see Eqs.~(\ref{result in conventional scenario 1}) 
and (\ref{result in conventional scenario 2})], 
ADBG predicts baryonic isocurvature perturbations such as 
\beq
 \mathcal{S}_{b \gamma} \equiv \frac{\delta Y_B}{Y_B} 
 \simeq n \cot \lmk n \theta_0 \rmk \delta \theta_0. 
 \label{S_b}
\eeq
Since the density perturbations of the CMB 
are predominantly adiabatic, 
the baryonic isocurvature perturbation is tightly constrained as~\cite{Ade:2015lrj} 
\beq
 \left\vert \mathcal{S}_{{\rm b} \gamma} \right\vert \lesssim 
5.0 \times 10^{-5}. 
\eeq
Therefore, this constraint puts an upper bound on the energy scale of inflation: 
\beq
 H_{\rm inf} \lesssim 5.3 \times 10^{14} \GEV \frac{\tan (n \theta_0)}{n} \frac{\abs{\phi}_{\rm inf}}{\Mpl}. 
\eeq
This can be rewritten as 
\beq
 H_{\rm inf} \lesssim 
 \left\{
 \bea{ll}
 1.6 \times 10^{13} \GEV \lmk \frac{\lambda}{2.6 \times 10^{-4}} \rmk^{-1} 
 &~~~~\text{for}~~ n = 4 
\\
 2.3 \times 10^{12} \GEV \lambda^{-1/3} 
 &~~~~\text{for}~~ n = 6, 
 \eea
 \right. 
 \label{isocurvature constraint} 
\eeq
where we use Eq.~(\ref{VEV}) 
and assume $\abs{c_H} = 1$ and $\tan (n \theta_0)=1$.

\section{\label{ADBG just after inf}Affleck-Dine baryogenesis just after inflation}

In this section, we explain a new scenario of ADBG 
where 
the AD field starts to oscillate around the origin of the potential 
just after the end of inflation. 
In general, this scenario is realized when the \Kahler potential is give by 
\beq
 K = \abs{\phi}^2 + \abs{S}^2 + \abs{\psi}^2 + \frac{c_1}{\Mpl^2} \abs{\phi}^2 \abs{S}^2 
 - \frac{c_2}{\Mpl^2} \abs{\phi}^2 \abs{\psi}^2, 
\label{Kahler}
\eeq
where $S$ is the field whose F-term drives inflation and $\psi$ is the field whose oscillation energy 
dominates the Universe after inflation. 
Here, we assume that the fields $S$ and $\psi$ are different fields, 
which is actually the case in hybrid and chaotic inflation models as shown in the subsequent sections.

During inflation, the AD field acquires the Hubble-induced mass 
via the F-term potential of the field $S$ as Eq.~(\ref{H-mass during inf}). 
After inflation ends, 
the Hubble-induced mass comes also from 
higher-dimensional kinetic interactions between $\phi$ and $\psi$ as Eq.~(\ref{H-mass after inf}). 
Therefore, the Hubble induced mass term for the AD field $\phi$ is given by 
\beq
 V_H &=& c_H H^2 (t) \abs{\phi}^2 \\
 c_H 
&=& 
\left\{ 
\bea{ll} 
- 3 (c_1 - 1) 
&~~~~\text{during \ inflation} \\
3 \lmk - (1-r ) c_1 + r c_2 + \frac{1}{2} \rmk 
&~~~~\text{after \ inflation}, \\
\eea
\right. 
\eeq
where $r$ ($0 \le r \le 1$) is the fraction of the energy density of $\psi$ to the total energy 
after inflation. 
Therefore 
the sign of the Hubble-induced mass term 
can change after inflation. 
If its sign continues to be negative after inflation, 
the conventional scenario of ADBG is realized as we explain in the previous section. 
In the rest of this paper, 
we consider the case that 
the coefficient is negative during inflation 
and is positive after inflation. 
In this case, 
the AD field starts to oscillate around the origin of the potential 
just after the end of inflation. 
In contrast to the conventional scenario of ADBG, 
the dynamics of its phase direction depends on inflation models, 
so that 
the resulting $B-L$ asymmetry depends on parameters 
in inflaton sector. 
In the subsequent sections, we consider hybrid and chaotic inflation models 
to investigate this scenario and calculate the amount of $B-L$ asymmetry. 
Before we investigate the detail of the dynamics of AD field, 
we explain its rough behaviour in this section.

In the above scenario, the dynamics of the AD field is determined by the potential of 
\beq
 V(\phi) = c_H H^2(t) \abs{\phi}^2 
 + \lambda^2 \frac{\abs{\phi}^{2n-2}}{\Mpl^{2n-6}} + V_A (\phi), 
\eeq
where 
$c_H < 0 $ during inflation 
and $c_H > 0$ after inflation. 
The A-term potential of $V_A$ depends on inflation models 
and is explicitly derived in the subsequent sections. 
The low-energy soft terms of Eq.~(\ref{A-term}) are irrelevant 
for the dynamics of the AD field. 
This makes our calculation simple and independent of low-energy SUSY models. 
In particular, the resulting $B-L$ asymmetry is independent of 
how SUSY breaking effect is mediated to the visible sector.

Since we consider the case that $c_H < 0 $ during inflation 
and $c_H > 0$ after inflation, 
the AD field starts to oscillate around the origin just after the end of inflation. 
At the same time, 
its phase direction is kicked by an A-term. 
The origin of A-term depends on inflation models 
and thus the resulting $B-L$ asymmetry does. 
Here we just write generated $B-L$ asymmetry as 
\beq
 \frac{a^3 (t)}{a^3 (t_{\rm osc})} n_{B-L} (t) \equiv q \epsilon H_{\rm osc} \abs{\phi}^2_{\rm osc}, 
\eeq
and derive $\epsilon$ in the subsequent sections. 
The resulting baryon-to-entropy ratio is thus written as 
\beq
 Y_b  
 &\simeq& 
 \left. \frac{8}{23} \frac{3 T_{\rm RH} n_{B-L}}{4 \rho_{\rm inf}} \right\vert_{\rm osc} \\
 &\simeq& 
 \frac{8}{23} \frac{\epsilon q T_{\rm RH}}{4 H_{\rm osc}} \lmk \frac{\phi_{\rm osc}}{\Mpl} \rmk^2. 
\eeq
This is the same with Eq.~(\ref{Y_b 1}) 
but $H_{\rm osc}$ is not given by Eqs.~(\ref{H_osc 1}) and (\ref{H_osc 1-2}). 
Since the AD field starts to oscillate just after the end of inflation in this scenario, 
$H_{\rm osc}$ is given by the Hubble parameter at the end of inflation. 
Here, let us emphasise 
differences from the conventional scenario of ADBG. 
The Hubble parameter at the time of beginning of oscillation $H_{\rm osc}$ 
is determined by the energy scale of inflation, 
not by 
either $m_\phi$ nor $T_{\rm RH}$ 
[see Eqs.~(\ref{H_osc 1}) and (\ref{H_osc 2})]. 
This is because the flat direction starts to oscillate just after the end of inflation 
due to the positive Hubble-induced mass term. 
In addition, $\phi_{\rm osc}$ depends only on $H_{\rm osc}$ and $\lambda$ via Eq.~(\ref{VEV2}). 
Therefore, 
the resulting $B-L$ asymmetry is independent of 
parameters in low-energy SUSY models, such as $m_\phi$ and $m_{3/2}$.

There are some advantages in this scenario. 
First, 
as we explain above, 
the resulting $B-L$ asymmetry is independent of 
the masses of the AD field and gravitino. 
The result is also independent of how SUSY breaking effect is mediated to the visible sector. 
Secondly, 
non-topological solitons, called Q-balls, may form and affects the cosmological scenario after the conventional scenario of ADBG~\cite{Coleman, Qsusy, EnMc, KuSh, KK}, 
while they do not form in our scenario. 
This makes the discussion much simpler. 
In particular, Q-balls usually form in gauge mediated SUSY breaking models 
after the conventional scenario of ADBG 
and they are sometimes problematic in cosmology due to their long lifetime~\cite{KK, Harigaya:2014tla}. 
Our scenario does not suffer from this problem. 
Thirdly, 
the thermal effect on the AD field can be neglected 
because 
the AD field starts to oscillate just after the end of inflation and before the thermal plasma grows sufficiently~\cite{Mukaida:2015ria}. 
This also makes calculations simpler. 
In particular, 
the thermal log potential can be neglected even for $L H_u$ flat direction. 
Finally, 
our results imply that 
ADBG works in broader range of parameter space. 
Since the sign of the Hubble-induced mass term 
cannot be determined by underlying physics, 
it is equally possible that the sign becomes positive after inflation. 
In addition, 
viable parameter regions for some parameters, e.g., the reheating temperature, 
are different from the ones in the conventional scenario of ADBG. 
These fact imply that the Affleck-Dine mechanism 
works well in more cases than expected in the literature.

\section{\label{hybrid}Hybrid inflation}

In this section, we consider our scenario of ADBG 
in the simplest hybrid inflation model~\cite{Copeland:1994vg, Dvali:1994ms} 
and calculate $B-L$ asymmetry. 
The superpotential in the inflaton sector is given by 
\beq
 W^{(\rm inf)} = \kappa S \lmk \psi \bar{\psi} - \mu^2 \rmk, 
\eeq
where $S$ is inflaton, and $\psi$ and $\bar{\psi}$ are waterfall fields. 
The F-term potentials are thus given as 
\beq
 \left. V_{\rm inf} \right\vert_{\rm tree} 
 = \kappa^2 \abs{\psi \bar{\psi} - \mu^2}^2 + \kappa^2 \abs{S}^2 \lmk \abs{\psi}^2 + \abs{\bar{\psi}}^2 \rmk. 
\eeq
The inflaton $S$ is assumed to have a large initial VEV 
so that the waterfall fields stay at the origin due to effective masses of $\kappa \la S \ra$. 
Then the F-term of $S$ is nonzero and drives inflation, 
where 
the energy scale of inflation is given by $3 H_{\rm inf}^2 \Mpl^2 \simeq \kappa^2 \mu^4$. 
The inflaton $S$ slowly rolls toward the origin due to the 1-loop Coleman-Weinberg potential: 
\beq
 \left. V_{\rm inf} \right\vert_{\rm 1-loop} 
 = 
 \frac{\kappa^4 \mu^4}{32 \pi^2}
 \lkk 
 \lmk x^2 + 1 \rmk^2 \ln \lmk x^2 + 1 \rmk 
+ \lmk x^2 - 1 \rmk^2 \ln \lmk x^2 - 1 \rmk  
- 2 x^4 \ln x^2 
- 3 
 \rkk, 
 \label{CW}
\eeq
where we define $x \equiv \abs{S}/\mu$. 
Inflation ends when its VEV decreases to the critical value of $S_{cr} \equiv \mu$. 
The Hubble parameter at the end of inflation is given by 
\beq
 H_{\rm osc} \simeq H_{\rm inf} \simeq \frac{\kappa \mu^2}{\sqrt{3} \Mpl}. 
\eeq
After that, the waterfall fields as well as the inflaton start to oscillate around the minimum of the potential 
and their oscillation energy dominates the Universe. 
Around the minimum of the potential, 
the masses of inflaton and waterfall fields are given by $\sqrt{2} \kappa \mu$.

Although the simplest hybrid inflation model 
predicts inconsistent spectral index with the observed value, 
some modifications can make it consistent. 
For example, 
we may introduce a higher dimensional \Kahler potential for the inflaton 
to write a small negative mass term, 
which can result in a consistent spectral index~\cite{BasteroGil:2006cm, Nakayama:2010xf}.
Since our discussion below is not affected at least quantitatively in this modification, 
we calculate $B-L$ asymmetry in the above simplest model.

\subsection{Dynamics of the AD field}

The inflaton $S$ is identified with the field $S$ in Eq.~(\ref{Kahler}) 
and 
the waterfall fields $\psi$ and $\bar{\psi}$ 
play a role of the field $\psi$ in Eq.~(\ref{Kahler}). 
Thus the coefficient of the Hubble-induced mass 
$c_H$ can change after inflation. 
In this subsection, we consider the dynamics of the AD field in the hybrid inflation model 
and calculate $B-L$ asymmetry.

Let us first consider the dynamics of the phase direction 
of the AD field. 
Using Eq.~(\ref{SUGRA potential}) with the total superpotential of $W^{(\rm AD)} + W^{(\rm inf)}$, 
we find that there is an A-term potential coming from 
\beq
 W^{(\rm inf)}_S K^{\bar{S} {\phi}} W^{(\rm AD)}_{\bar{\phi}} 
 + K_\phi W^{(\rm inf)} \lmk W^{(\rm AD)}_\phi \rmk^* 
 + K_S W^{(\rm AD)} \lmk W^{(\rm inf)}_S \rmk^* 
 - 3 W^{(\rm inf)} \lmk W^{(\rm AD)} \rmk^* + {\rm c.c.} 
\eeq
The A-term is written as 
\beq
 V_A 
 &=&
  - \lmk 1- c_1 - \frac{2}{n} \rmk \frac{\kappa \mu^2 \lambda}{\Mpl^{n-1}} S^* \phi^n + {\rm c.c.} \\
 &=&
 - a \frac{H^2_{\rm inf}}{\Mpl} \abs{S} \abs{\phi}^2 \cos \lmk \theta_S - n \theta_\phi \rmk, 
 \label{V_A hybrid}\\
 a &\equiv&  - 2 \lmk c_1 - 1 + \frac{2}{n} \rmk \sqrt{\frac{3 \abs{c_H}}{n-1} }, 
\eeq
where $\theta_S$ and $\theta_\phi$ are the complex phases of the fields $S$ and $\phi$, respectively. 
We use Eq.~(\ref{VEV}) and $H_{\rm inf}^2 = \kappa^2 \mu^4 / 3 \Mpl^2$ in the second line. 
This is a linear term of the inflaton $S$, 
so that 
the slope of the potential should not be larger than that of the Coleman-Weinberg potential~\cite{
Buchmuller:2000zm, Nakayama:2010xf, Buchmuller:2014epa}. 
Otherwise the inflaton cannot reach the critical VEV and inflation cannot terminate 
unless we allow a fine-tuning on the initial phase of inflaton. 
Referring to Ref.~\cite{Buchmuller:2014epa}, 
we introduce a parameter to describe the relative importance of the 
two contributions to the slope of the potential: 
\beq
 \xi &\equiv& 
 \frac{1}{2} \lmk 1- c_1 - \frac{2}{n} \rmk 
 \frac{16 \pi^2}{\kappa^3 \ln 2} \frac{\la \abs{\phi} \ra^n}{\mu \Mpl^{n-1}} \\
 &\simeq& \frac{8 \pi^2 a}{3 \ln 2} \frac{\mu \la \abs{\phi} \ra^2}{\kappa^2 \Mpl^3}, 
 \label{constraint3}
\eeq
which should be smaller than unify so that 
the inflaton can roll towards the critical value without the fine-tuning.%
\footnote{
When the VEV of the AD field is so large that 
the parameter $\xi$ becomes of order unity (but below unity), 
the A-term of Eq.~(\ref{V_A hybrid}) affects inflaton dynamics. 
As a result, 
the spectral index can be consistent with the observed value~\cite{Yamada:2015rza}. 
}

In the above minimal setup, there is no other term than Eq.~(\ref{V_A hybrid}) 
that affects the dynamics of the phase directions. 
Therefore, there is only one massive phase during inflation. 
For simplicity, let us assume that the inflaton and the AD field have 
approximately constant VEVs and $(\theta_S - n \theta_\phi) \ll 1$. 
In this case, 
the unitary matrix to diagonalise the squared mass matrix for the phase directions 
is given by 
\beq
\frac{1}{\sqrt{n^2 \abs{S}^2 + \abs{\phi}^2}} 
 \lmk 
 \bea{ll} 
 \abs{\phi}~~~ - n \abs{S}
 \vspace{0.2cm}\\
 n \abs{S} ~~~\abs{\phi} 
 \eea
 \rmk, 
\label{unitary matrix}
\eeq
in the $( \abs{S} \theta_S / \sqrt{2} , \abs{\phi} \theta_\phi / \sqrt{2} )^T$ basis. 
Thus, the massive direction denoted by $f_m \theta_m$ can be written as 
\beq
 f_m \theta_m = \frac{\sqrt{2}  \abs{S} \abs{\phi}}{\sqrt{n^2 \abs{S}^2 + \abs{\phi}^2}} \lmk \theta_S - n \theta_\phi \rmk, 
\eeq
and its mass $m_{\theta_m}$ is given by 
\beq
 m_{\theta_m} = \sqrt{\frac{a H^2}{2} \frac{\abs{\phi}}{\Mpl} \lmk \frac{\abs{\phi}}{\abs{S}} + n^2\frac{\abs{S}}{\abs{\phi}} \rmk}.
\eeq
If the curvature of the phase direction is larger than the Hubble parameter during inflation, 
it stays at the minimum of the A-term, i.e., $\theta_m = 0$, 
and the phase direction cannot be kicked in the complex plane after inflation. 
In this case, $B-L$ asymmetry cannot be generated. 
Thus, we require $m_{\theta_m} \ll H$, 
which can be rewritten as 
\beq
 a \abs{\phi}^2 &\ll& \abs{S} \Mpl, 
 \label{constraint1}\\
 a n^2 \abs{S} &\ll& \Mpl, 
 \label{constraint2}
\eeq
in order that the phase direction can stay at a different phase from the minimum 
due to the Hubble friction effect. 
We denote the initial phase as $\theta_m^{\rm ini}$.

After inflation ends, 
the AD field acquires a positive Hubble-induced mass term 
and starts to oscillate around the origin of the potential. 
At the same time, 
the massive phase direction is kicked by the above A-term. 
Since the radial direction decreases with time due to the Hubble expansion, 
the A-term is relevant just after the beginning of oscillation. 
Thus we can estimate the angular velocity of massive phase direction 
such as 
\beq
 \dot{\theta}_m \approx \frac{m_{\theta_m}^2}{H} \theta_m^{\rm ini}, 
\eeq
[see Eq.~(\ref{EOM for phase direction})]. 
Using the inverse of the unitary matrix of Eq.~(\ref{unitary matrix}), 
we obtain the angular velocity of the phase of the AD field such as 
\beq
 \dot{\theta}_\phi 
 &=& 
 \frac{ - n \abs{S}}{\sqrt{n^2 \abs{S}^2 + \abs{\phi}^2}}
 \frac{f_m}{\sqrt{2} \abs{\phi}} \dot{\theta}_m 
 \\
 &\approx& 
 \frac{m_{\theta_m}^2}{H} 
 \frac{ - n \abs{S}}{\sqrt{n^2 \abs{S}^2 + \abs{\phi}^2}} 
 \frac{f_m \theta_m^{\rm ini}}{\sqrt{2} \abs{\phi}}  
 \\
 &=& 
 \frac{m_{\theta_m}^2}{H} 
 \frac{ - n \abs{S}}{\sqrt{n^2 \abs{S}^2 + \abs{\phi}^2}}
 \frac{ \abs{S}}{\sqrt{n^2 \abs{S}^2 + \abs{\phi}^2}} \lmk \theta_S - n \theta_\phi \rmk^{\rm ini} 
\\
 &=& - \frac{a n}{2} \frac{\abs{S}}{\Mpl} H \lmk \theta_S - n \theta_\phi \rmk^{\rm ini}. 
\eeq
Thus we obtain 
\beq
  \frac{a^3(t)}{a^3 (t_{\rm osc})} n_{B-L} (t) 
  &=& \left. 2 \dot{\theta}_\phi \abs{\phi}^2 \right\vert_{\rm osc} 
  \\
 &\equiv& 
 \epsilon q H_{\rm osc} \phi^2 \\
 \epsilon &\equiv& \tilde{\epsilon} \frac{S_{\rm cr}}{\Mpl} 
 \label{epsilon}\\
 \tilde{\epsilon}
 &\simeq& 
 (0.1-0.2)
 a n 
 \sin \lmk n \theta_\phi - \theta_S \rmk_{\rm osc}, 
 \label{epsilon-factor}
\eeq
where we define $\tilde{\epsilon}$ which is expected to be of order unity. 
The numerical factor of $(0.1-0.2)$ is determined from our numerical calculations explained below. 
Note that 
the resulting ellipticity parameter $\epsilon$ 
is consistent with a naive estimation of 
$\epsilon \sim V_A' / \phi H_{\rm osc}^2$.%
\footnote{
We implicitly assume that $(S_{cr} / \Mpl ) \gtrsim m_{3/2} / H_{\rm osc}$ 
so that we can neglect an A-term of $m_{3/2} W_\phi$ [see Eq.~(\ref{A-term})]. 
Otherwise $\epsilon$ 
may be of order $m_{3/2} / H_{\rm osc}$. 
}
The ellipticity parameter $\epsilon$, which describes the efficiency of baryogenesis, 
is much smaller than unity 
because of the condition of Eq.~(\ref{constraint2}). 
This is because the phase direction of the AD field is kicked by the A-term 
that is suppressed by the VEV of the inflaton.

After the oscillations begins, 
the amplitude of the radial direction of the inflaton $S$ decreases with time as $\abs{S} \propto a^{-3/2}$. 
That of the AD field does as $\abs{\phi} \propto a^{-3/4}$ 
so that its number density ($H(t) \abs{\phi}^2/2$) decreases as $\propto a^{-3}$. 
Since the A-term, i.e., the $B-L$ number violating interaction, is a higher dimensional term, 
it is turned off soon after the AD field starts to oscillate after inflation. 
The generated $B-L$ asymmetry is then conserved in a comoving volume 
and thus $n_{B-L} \propto a^{-3}$ for $t > t_{\rm osc}$.

\begin{figure}[t]
\centering 
\includegraphics[width=.45\textwidth, bb=0 0 360 240
]{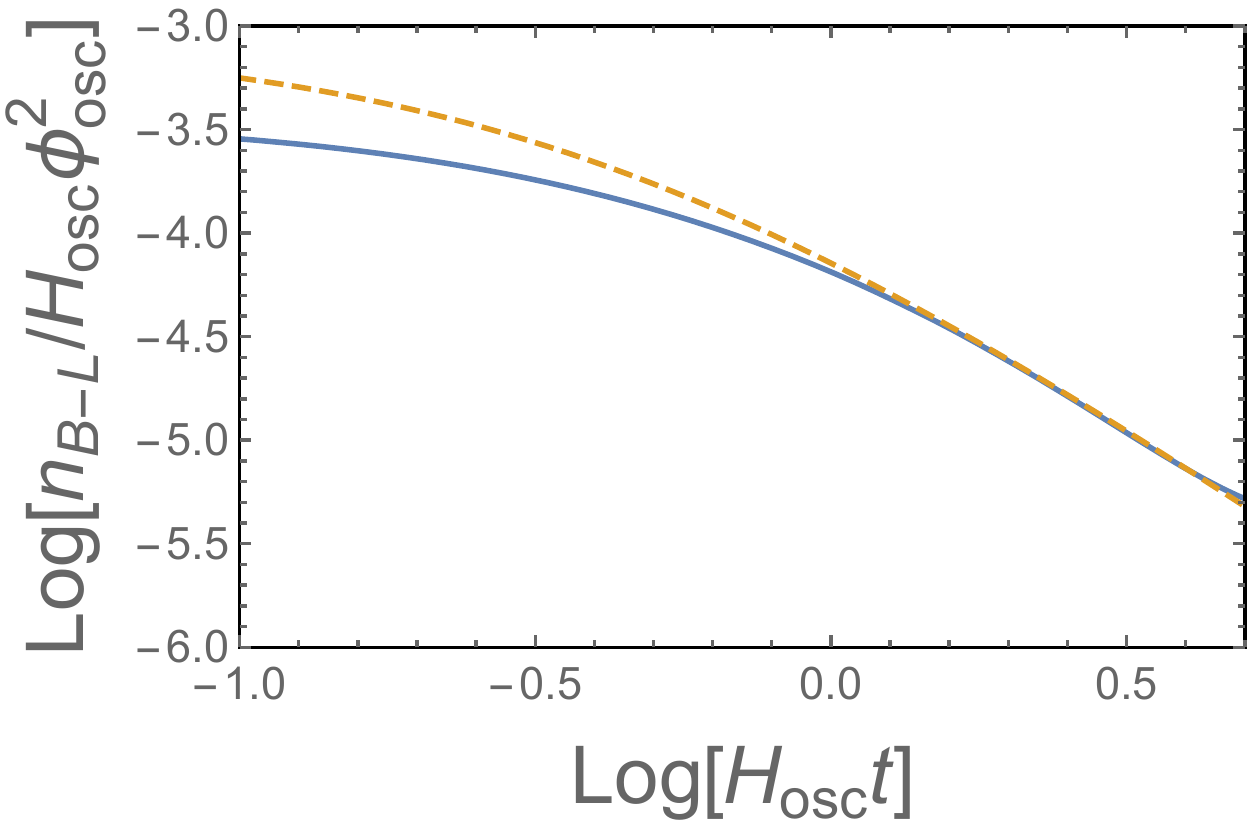} 
\caption{\small
Evolution plot for $B-L$ number after hybrid inflation. 
The dashed curve is our prediction of Eq.~(\ref{epsilon-factor}) with a numerical factor of $0.2$. 
We assume $\lambda=1$, $n=6$, $c_H = -1$, $c_2=0$, $\kappa = 0.05$, $\mu = 0.001$, and $\theta_\phi^{\rm ini} = 0.01$. 
}
  \label{fig3}
\end{figure}

We have numerically solved the equations of motion 
together with the Friedmann equation, 
where the waterfall fields are collectively described by a real scalar field $\tilde{\psi}$ 
such as $\psi = \bar{\psi} \equiv \tilde{\psi}/\sqrt{2}$. 
We assume 
$\abs{S}^2 / \Mpl^2$, $\abs{\phi}^2 / \Mpl^2$, $\tilde{\psi}^2 / \Mpl^2 \ll 1$ 
and 
take into account next-to-leading order terms in terms of them. 
We use the full kinetic terms for $S$ and $\phi$ [see Eq.~(\ref{kinetic term})], 
while we assume a canonical one for $\psi$ for simplicity. 
One of the results is shown in Fig.~\ref{fig3}, 
where the generated $B-L$ asymmetry is consistent with Eq.~(\ref{epsilon}). 
Taking parameters such as $n=4, 6$, $\kappa = 0.02-0.5$, $\mu = 0.0004 - 0.02$, 
$\lambda = 0.01 - 100$, and $\theta_\phi^{\rm ini} = 0.001 - 0.1$, 
we confirm the above parameter dependences 
and obtain the numerical uncertainty of $(0.1-0.2)$ in Eq.~(\ref{epsilon-factor}). 
We assume $c_H = -1$ and $c_2=0$ in our calculations, 
but we check that nonzero values of $c_2$ ($= \mathcal{O}(1)$ and $\ge 0$) does not change our results 
even quantitatively.

\subsection{Baryon asymmetry}

The AD field starts to oscillate just after inflation and generate $B-L$ asymmetry. 
The oscillating AD field decays and dissipates into radiation~\cite{Mukaida:2012qn} 
and the sphaleron effect relates 
the $B-L$ asymmetry to the baryon asymmetry~\cite{Kuzmin:1985mm, Fukugita:1986hr}. 
Using Eq.~(\ref{epsilon}), 
we can calculate the baryon-to-entropy ratio $Y_b$ such as 
\beq
 Y_b  
 &\simeq& 
 \frac{8}{23} \frac{\epsilon q T_{\rm RH}}{4 H_{\rm osc}} \lmk \frac{\phi_{\rm osc}}{\Mpl} \rmk^2 \\
 &\simeq& 
 \left\{
 \bea{ll}
 0.05 \sqrt{\abs{c_H}} q \frac{\epsilon}{\lambda} \frac{T_{\rm RH}}{\Mpl} 
 ~~~~\text{for}~~ n = 4 
  \vspace{0.2cm}\\
 0.06 \abs{c_H}^{1/4} q \frac{\epsilon}{\lambda^{1/2}} \frac{T_{\rm RH}}{\sqrt{H_{\rm osc} \Mpl}} 
 ~~~~\text{for}~~ n = 6, 
 \eea
 \right. 
 \label{Y_b}
\eeq
Since 
$\epsilon \equiv \tilde{\epsilon} S_{\rm cr} / \Mpl$, $S_{\rm cr} = \mu$, and 
$H_{\rm osc}^2 \simeq \kappa^2 \mu^4 / (3 \Mpl^2)$, 
this is rewritten as 
\beq
Y_b 
 &\simeq& 
 \left\{
 \bea{ll}
 0.05 
 \frac{\mu T_{\rm RH}}{\lambda \Mpl^2} 
 ~~~~\text{for}~~ n = 4 
 \vspace{0.2cm}\\
 0.08 
 \frac{T_{\rm RH}}{ \sqrt{\kappa \lambda} \Mpl} 
 ~~~~\text{for}~~ n = 6, 
 \eea
 \right.
\eeq
where we assume $\abs{c_H} = 1$, $q=1$, and $\tilde{\epsilon} = 1$. 
For typical parameters, 
it is given by 
\beq
Y_b 
 &\simeq& 
 \left\{
 \bea{ll}
  9 \times 10^{-11} 
  \lmk \frac{\mu}{10^{15}  \GEV} \rmk 
  \lmk \frac{T_{\rm RH}}{10^{9}  \GEV} \rmk 
  \lmk \frac{\lambda }{10^{-4}} \rmk^{-1} 
 &~~~~\text{for}~~ n = 4 
 \vspace{0.2cm}\\
 1 \times 10^{-10} \lambda^{-1/2} 
   \lmk \frac{\kappa}{10^{-3} } \rmk^{-1/2} 
  \lmk \frac{T_{\rm RH}}{10^{7}  \GEV} \rmk 
 &~~~~\text{for}~~ n = 6, 
 \eea
 \right.
 \label{result in hybrid}
\eeq
We check that 
the constraints of Eqs.~(\ref{constraint1}) and (\ref{constraint2}) 
and $\xi \le 1$ [see Eq.~(\ref{constraint3})] 
are satisfied 
for the above reference parameters. 
Thus, we can explain 
the observed baryon asymmetry of 
$Y_b^{\rm obs} \simeq 8.7 \times 10^{-11}$~\cite{pdg} in this scenario.

Since a linear combination of phase directions is massless during inflation, 
our scenario predicts nonzero baryonic isocurvature fluctuations 
like the case in Sec.~\ref{isocurvature}. 
However, 
the energy scale of hybrid inflation can be lower than the constraint of Eq.~(\ref{isocurvature constraint}). 
In fact, 
for the above reference parameters, 
our scenario is consistent with the present upper bound on the 
isocurvature mode.

\subsection{Reheating temperature}

As we can see in Eq.~(\ref{result in hybrid}), 
the resulting baryon asymmetry depends on reheating temperature $T_{\rm RH}$. 
To determine it, let us consider the decay of inflaton. 
There is a lower bound on the reheating temperature 
because the inflaton decays into the MSSM particles via supergravity effects. 
The decay rate is calculated as 
\beq
 \Gamma_{\rm inf}^{\rm SUGRA} = \frac{3}{128 \pi^3} \abs{y_t}^2 
 \lmk \frac{\mu}{\Mpl} \rmk^2 \frac{m_{\rm inf}^3}{\Mpl^2}, 
\eeq
where $m_{\rm inf} = \sqrt{2} \kappa \mu$ is the inflaton mass 
and $y_t$ is the top Yukawa coupling constant. 
The lower bound on the reheating temperature is thus given by~\cite{Nakamura:2006uc} 
\beq
 T_{\rm RH}^{(\rm min)} \simeq 3 \times 10^3 \GEV \abs{y_t} \lmk \frac{\mu}{10^{15} \GEV} \rmk 
 \lmk \frac{m_{\rm inf}}{10^{12} \GEV} \rmk^{3/2}. 
\eeq
If there is an interaction between the inflaton and Higgs fields such as 
\beq
 W \supset y \phi H_u H_d, 
\eeq
then the inflaton decay rate and the reheating temperature 
are estimated as 
\beq
 \Gamma_{\rm inf} &=& \frac{y^2}{4 \pi} m_\phi 
 \\ 
 T_{\rm RH} &\simeq& 
 2 \times 10^{10} \GEV 
 \lmk \frac{y}{10^{-4}} \rmk 
 \lmk \frac{m_{\rm inf}}{10^{12} \GEV} \rmk^{1/2}. 
 \label{T_RH in hybrid}
\eeq
Note that 
the coupling constant $y$ should be smaller than $\kappa$ 
so as not to affect the Coleman-Weinberg potential of Eq.~(\ref{CW}). 
Thus the reheating temperature cannot be higher than 
that of Eq.~(\ref{T_RH in hybrid}) with $y \approx \kappa$.

We have to take into account the constraint on $T_{\rm RH}$ 
from gravitino overproduction problems. 
The inflaton decays also into gravitinos via supergravity effects. 
Its production rate is given by~\cite{Nakamura:2006uc} 
\beq
 \Gamma_{3/2} \simeq \frac{1}{96 \pi} \lmk \frac{\mu}{\Mpl} \rmk^2 \frac{m_{\rm inf}^3 }{\Mpl^2}. 
\eeq
The resulting gravitino-to-entropy ratio from this contribution is given by
\beq
 Y_{3/2}^{\rm (decay)} 
 \simeq \frac{3}{2} \lmk \frac{90}{g_* \pi^2} \rmk^{1/2} \frac{\Gamma_{3/2} \Mpl}{m_{\rm inf} T_{\rm RH}}. 
\eeq
Gravitinos are also produced from scatterings in the thermal plasma after reheating completes. 
Its abundance is given by~\cite{Bolz:2000fu, Pradler:2006qh, Buchmuller:2011mw}
\beq
 Y_{3/2}^{\rm (thermal)} 
 \simeq 
 0.26 
 \frac{\rho_c}{m_{3/2} s_0} \lmk \frac{T_{\rm RH}}{10^{10} \GEV} \rmk 
 \lkk 0.13 
 \lmk \frac{m_{3/2}}{100 \GEV} \rmk 
 + 
 \lmk \frac{100 \GEV}{m_{3/2}} \rmk 
 \lmk \frac{m_{\tilde{g}}}{1 \TEV} \rmk^2 
 \rkk, 
\eeq
where $s_0$ ($\simeq 2.9 \times 10^3 \cm^{-3}$) and $\rho_c$ ($\simeq 1.052 \times 10^{-5} h^2 
\GEV / \cm^3$) are the present entropy density and critical energy density, respectively. 
The parameter $m_{\tilde{g}}$ is gluino mass and $h$ is the present 
Hubble parameter in the unit of $100 \km \, s^{-1} \Mpc^{-1}$. 
Stringent bounds on the reheating temperature 
are obtained 
when we assume that the gravitino is the lightest SUSY particle (LSP) 
and is stable. 
In this case, 
its abundance should not exceed the observed DM abundance: 
\beq
 m_{3/2} \lmk Y_{3/2}^{\rm (decay)} + Y_{3/2}^{\rm (thermal)} \rmk 
 \le \frac{\rho_c}{s_0} \Omega_{\rm DM} 
 \simeq 0.4 \EV, 
 \label{gravitino problem2}
\eeq
where $\Omega_{\rm DM} h^2$ ($\simeq 0.12$) is the 
DM relic density.%
\footnote{
If the gravitino mass is about $1 \TEV$ and it is unstable, 
its decay products interact with the light elements 
and destroy them at the time of BBN epoch. 
Then the gravitino abundance is bounded above by about four order of magnitude severer than 
the bound of Eq.~(\ref{gravitino problem2})~\cite{Kawasaki:1999na}. 
}
For example, in the case of $m_{3/2} = 100 \GEV$, 
the reheating temperature is bounded such as 
\beq
 2 \times 10^{7} \GEV 
  \lmk \frac{\mu}{10^{15} \GEV} \rmk^2 
 \lmk \frac{m_{\rm inf}}{10^{12} \GEV} \rmk^2 
 \lesssim T_{\rm RH} \lesssim 9 \times 10^9 \GEV, 
\eeq
where we use $h \simeq 0.67$. 
We can see that the reference parameters used in Eq.~(\ref{result in hybrid}) are 
consistent with this bound.

Note that 
for the case of $n=4$, 
the coupling constant in the superpotential of the AD field 
cannot be much larger than $10^{-4}$ 
because of the upper bound on the reheating temperature. 
For the case of $n=6$, 
we can naturally explain 
the observed baryon asymmetry for $\lambda = \mathcal{O}(1)$ 
with a reheating temperature consistent with the gravitino problem. 
This is in contrast to the result in the conventional scenario of ADBG 
[see Eq.~(\ref{Y_b conventional})], 
where an extremely large value of $\lambda$ is required to be consistent with 
the lower bound on reheating temperature. 
In the case of such a large value of $\lambda$, 
the thermal log potential has to be taken into account even for $n=6$.

\section{\label{chaotic}Chaotic inflation}

In this section, we consider our scenario of ADBG 
in a chaotic inflation model with a shift symmetry in supergravity~\cite{Kawasaki:2000yn, Kallosh:2010ug}. 
The inflaton $I$ has a shift symmetry in the \Kahler potential 
and the minimal \Kahler potential is written as 
\beq
 K_{\rm inf} = 
  c_0 \Mpl \lmk I + I^* \rmk + 
 \frac{1}{2} \lmk I + I^* \rmk^2 + \abs{X}^2 
 - \frac{c_3}{4} \frac{ \abs{X}^4}{\Mpl^2}, 
\eeq
where $X$ is a stabiliser field. 
Note that $c_0$ is an order parameter 
of $Z_2$ symmetry, under which the fields $I$ and $X$ are odd, 
so that we take $c_0$ as a free parameter that may be smaller than unity. 
We include the $\abs{X}^4$ term in the \Kahler potential, 
which cannot be suppressed by any symmetries. 
The other higher dimensional terms do not change our discussion qualitatively, 
so that we neglect them in the following analysis.

\begin{figure}[t]
\centering 
\includegraphics[width=.25\textwidth, bb=0 0 309 541
]{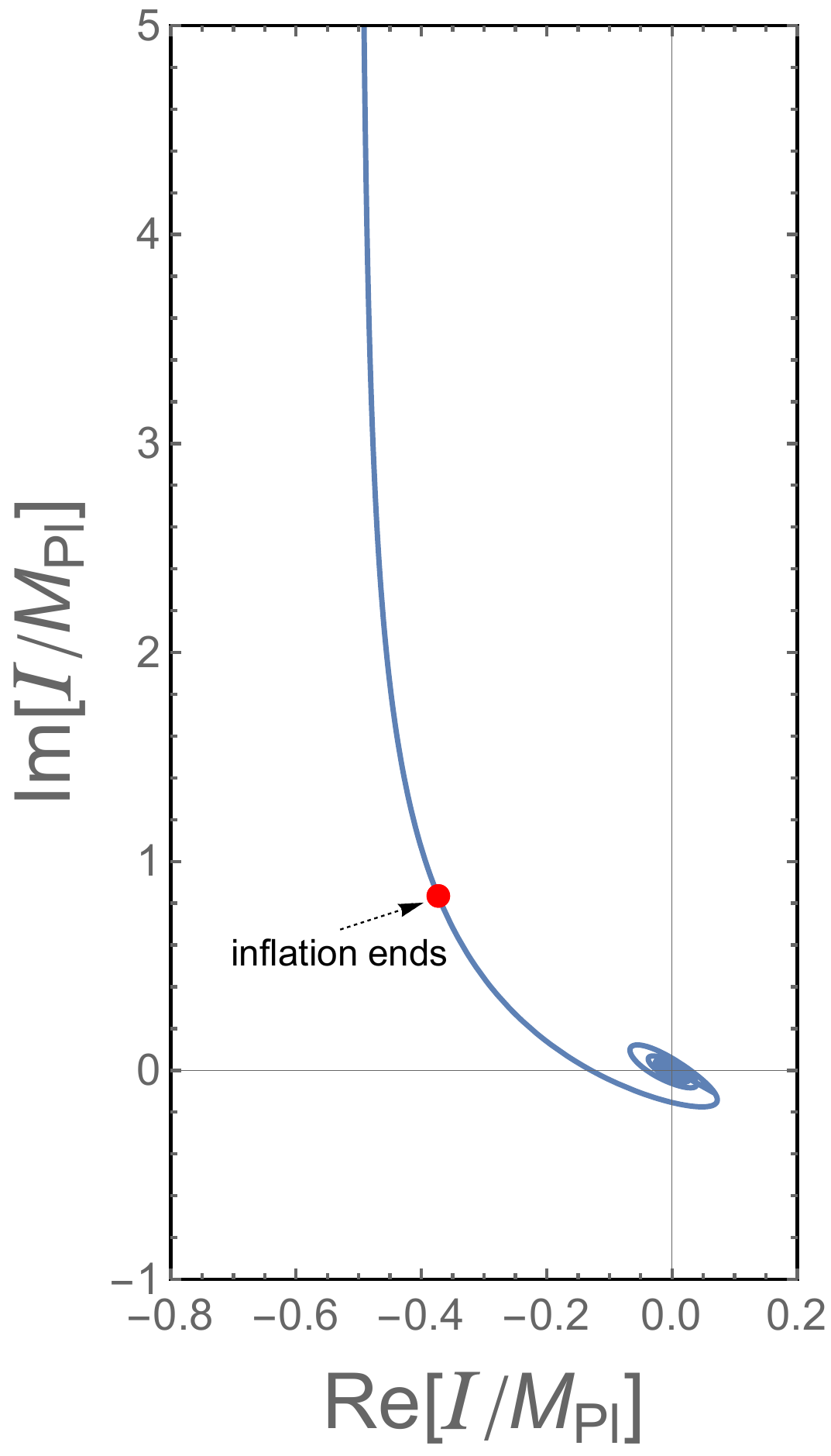} 
\caption{\small
Dynamics of the field $I$ in the complex plane in the chaotic inflation model. 
We set $c_0 = 1$. 
The field $I$ slowly rolls along the line of ${\rm Re} [ I ] = - c_0 / 2$ during inflation. 
After it reaches the red point, 
inflation ends and 
it starts to oscillate and rotate around the origin. 
}
  \label{fig4}
\end{figure}

To realize chaotic inflation in a quadratic potential, 
the superpotential is assumed to break the shift symmetry such as 
\beq
 W_{\rm inf} = m_{\rm inf} I X, 
\eeq
where $m_{\rm inf}$ is inflaton mass. 
The field $I$ 
has a quadratic potential from the F-term of $X$. 
Its imaginary component 
can have a larger VEV than the Planck scale thanks to the shift symmetry in the \Kahler potential 
and is identified with inflaton. 
The real component of $I$ obtains a Hubble-induced mass 
and stays at a VEV of ${\rm Re} [ I ] = - c_0 / 2$~\cite{Kawasaki:2000ws}. 
When the VEV of the inflaton decreases down to the Planck scale, 
the real component of $I$ as well as the inflaton start to oscillate around the origin of the potential 
and inflation ends. 
The dynamics is illustrated in Fig.~\ref{fig4}, 
where we numerically solve the equation of motion of the field $I$ 
and plot its trajectory for the case of $c_0 = 1$. 
The field $I$ slowly rolls along the imaginary axis during inflation, 
where ${\rm Re} [ I ] = - c_0 / 2$ is approximately satisfied. 
After it reaches the red point, 
inflation ends and 
it starts to oscillate and rotate around the origin. 
The Hubble parameter at the end of inflation is given by 
\beq
 H_{\rm osc} \simeq \frac{m_{\rm inf} }{ \sqrt{3}}. 
\eeq

The stabiliser field $X$ obtains a Hubble-induced mass 
via the higher dimensional \Kahler potential such as 
\beq
 V \supset c_3 m^2 \frac{ \abs{I}^2}{\Mpl^2} \abs{X}^2 
 \simeq 3 c_3 H^2 \abs{X}^2. 
 \label{Hubble induced mass of X}
\eeq
This implies that the dynamics of $X$ is qualitatively different 
from the case with $c_3 = 0$. 
We should include them 
because the higher dimensional \Kahler potential cannot be suppressed by any symmetries, 
To realize chaotic inflation, 
we assume $c_3 > 0 $. 
Then the field $X$ stays at the origin. 
However, 
when we take into account of the backreaction of the AD field, 
$X$ obtains a small VEV as shown in the next subsection.

\subsection{Dynamics of the AD field}

Taking into account the AD field, 
we consider the \Kahler potential of 
\beq
 K = 
 K_{\rm inf} 
 + \abs{\phi}^2 
 + c_1 \abs{X}^2 \abs{\phi}^2 
 - \frac{c_2}{2} \lmk I + I^* \rmk^2 \abs{\phi}^2. 
\eeq
Although we introduce a shift symmetry for the field $I$, 
the fields $X$ and $I$
basically correspond to 
the fields $S$ and $\psi$ in Eq.~(\ref{Kahler}), respectively. 
The AD field acquires 
the Hubble-induced mass term from the F-term of $X$ during inflation. 
After inflation ends, 
the Hubble-induced mass term 
partially comes from kinetic interactions. 
In fact, 
the \Kahler potential of $- c_2 /2 ( I+ I ^*)^2 \abs{\phi}^2$ 
induces a kinetic interaction of 
\beq
 \mathcal{L} \supset - c_2 \frac{1}{\Mpl^2} \abs{\phi}^2 \abs{\del_\mu I}^2. 
\eeq
We obtain the effective Hubble-induced mass term 
of $(3 c_2/2)  H^2 (t) \abs{\phi}^2$ from this kinetic interaction. 
To sum up, 
the Hubble-induced mass term is given by 
\beq
 V_H
 = 
 c_H H^2(t) \abs{\phi}^2 \\
 c_H 
 &=& 
 \left\{ 
 \bea{ll}
 - 3 (c_1 -1 ) 
 &~~~~\text{during \ inflation} \\
 \frac{3}{2} \lmk c_2 - c_1 + 1 \rmk 
 &~~~~\text{after \ inflation}, 
 \eea
 \right.
\eeq
where the other terms than the one proportional to $c_2$ come from the potential energy. 
Thus we can consider the case that 
the coefficient $c_H$ is negative during inflation 
and is positive after inflation. 

There is also an A-term such as 
\beq
 V_A 
 &=& 
 \frac{1}{n} \lmk n(1-c_1) - 2 \rmk 
 \frac{\lambda m_{\rm inf}}{\Mpl^{n-1}} I X (\phi^*)^n + {\rm c.c.} \\
 &=& \frac{2}{n} \lmk n(1-c_1) - 2 \rmk 
 \frac{\lambda m_{\rm inf}}{\Mpl^{n-1}} \abs{I} \abs{X} \abs{\phi}^n \cos \lmk \theta_I + \theta_X - n \theta_\phi \rmk \\
 &\simeq& 
 - a 
 H^2(t) \frac{ \abs{X}}{\Mpl} \abs{\phi}^2 \cos \lmk \theta_I + \theta_X - n \theta_\phi \rmk, 
 \label{A-term in chaotic inflation}
\eeq
where 
we use Eq.~(\ref{VEV}) and $H (t) \simeq m_{\rm inf} \abs{I} / \sqrt{3} \Mpl$ in the last line 
and $\theta_I$, $\theta_X$, and $\theta_\phi$ are the complex phases 
of the fields $I$, $X$, and $\phi$, respectively. 
The coefficient $a$ is given by 
\beq
 a = 
 2 \sqrt{\frac{3\abs{c_H}}{n-1}} \lmk c_1 - 1 + \frac{2}{n} \rmk. 
\eeq
The A-term can be regarded as a linear term for $X$. 
Since the field $X$ has a positive Hubble-induced mass term of Eq.~(\ref{Hubble induced mass of X}), 
it stays at the following minimum during inflation: 
\beq
 \la \abs{X} \ra 
 \simeq 
 \frac{a}{6 c_3} \frac{1}{\Mpl} \abs{\phi}^2. 
\eeq
A linear combination of the phase directions has a mass 
of order the Hubble parameter due to the A-term, 
so that it stays at the following minimum during inflation: 
\beq
 \la \theta_X - n \theta_\phi \ra = - \la \theta_I \ra \simeq - {\rm sign} [c_0 ] \frac{\pi}{2}, 
\eeq
where we use ${\rm Re} [ I ] \ll {\rm Im} [I]$ during inflation. 

\begin{figure}[t]
\centering 
\includegraphics[width=.45\textwidth, bb=0 0 450 302
]{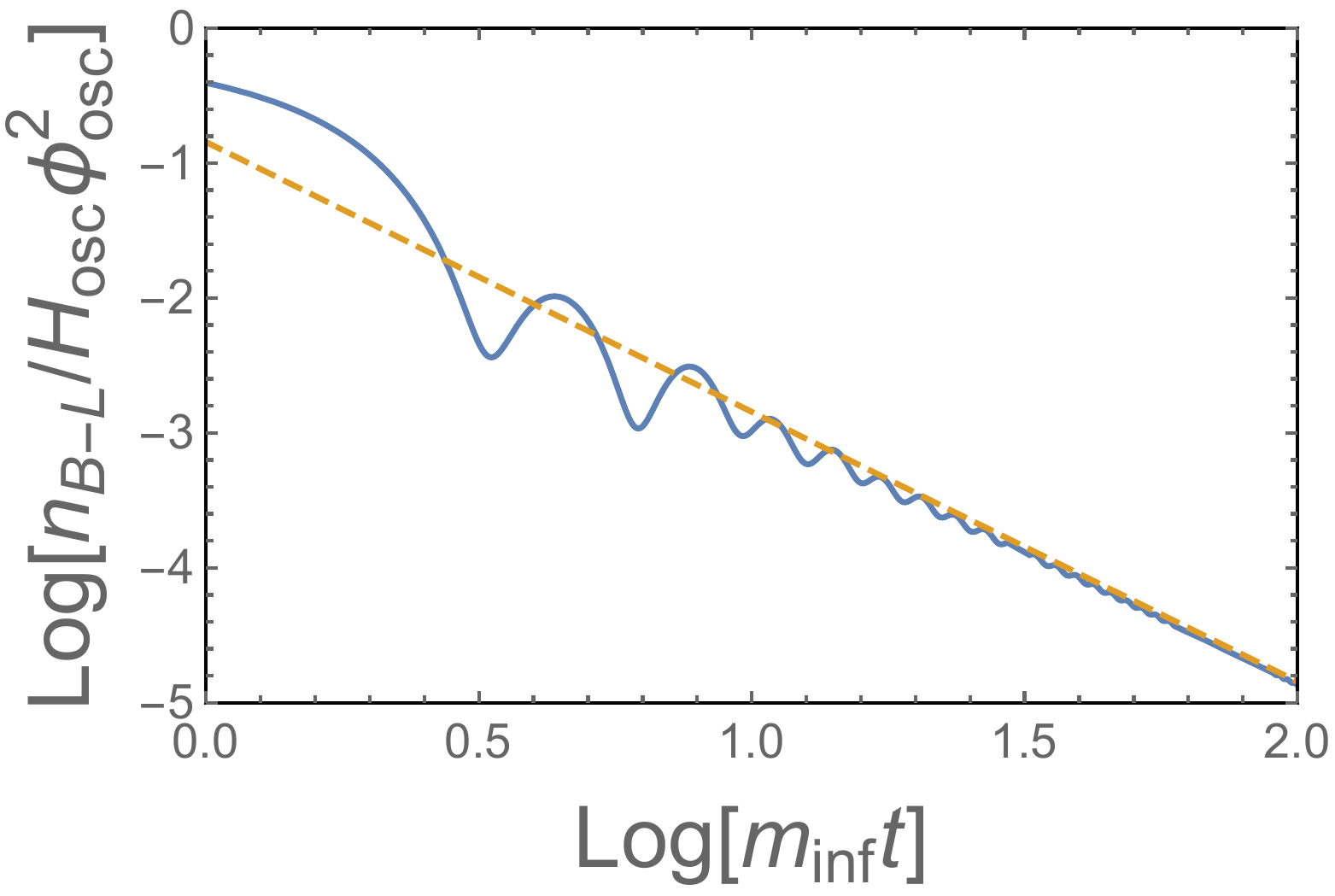} 
\caption{\small
Evolution plot for $B-L$ number density in our scenario of ADBG 
in the chaotic inflation model. 
The dashed curve is our prediction of Eq.~(\ref{numerical factor in chaotic inf}) with a numerical factor of $0.01$. 
We take $\lambda=1$, $n=6$, $c_0 = 0.5$, $c_1 = 2$, $\abs{c_2} = 1$, and $c_3 = -1$. 
}
  \label{fig5}
\end{figure}

After inflation ends, 
the field $I$
starts to rotate in the phase space as shown in Fig.~\ref{fig1} 
and its phase $\theta_I$ has a nonzero velocity. 
This implies that a linear combination of the phases $\theta_X$ and $\theta_\phi$ 
obtains a nonzero velocity to follow its potential minimum. 
Since the A-term contains the phase direction of the inflaton, 
the whole dynamics is difficult to imagine. 
In fact, 
one may estimate $\epsilon \approx a \abs{X}_{\rm osc}/ \Mpl$ 
like the case in the hybrid inflation model considered in the previous section [see Eq.~(\ref{epsilon})], 
but we find this estimation wrong. 
We perform numerical calculations 
to solve the equations of motion for the complex scalar fields $S$, $X$, and $\phi$. 
We use the full supergravity potential for $S$, $X$, and $\phi$. 
The kinetic interactions are simplified 
such that $S$ and $X$ have canonical kinetic terms for simplicity. 
We take into account the kinetic interactions for $\phi$ associated with $c_2$, 
which is needed to change the sign of its Hubble-induced mass term. 
The parameters are taken in the intervals of $\lambda = 10^{-3} - 10^4$ and $c_0 = 10^{-5} - 1$ 
for $n=4$ and $6$. 
The $\mathcal{O}(1)$ coefficients in the \Kahler potential 
are assumed to be $c_1=2$, $c_2 = 1$, and $c_3=1$. 
From our numerical calculations, we obtain the following results: 
\beq
  \frac{a^3(t)}{a^3 (t_{\rm osc})} n_{B-L} (t) 
 &\equiv& 
 \epsilon q H_{\rm osc} \phi_{\rm osc}^2 
 \\
 \epsilon 
 &\equiv& \tilde{\epsilon} c_0 
 \label{result in chaotic inf}\\ 
 \tilde{\epsilon} 
 &\simeq&
 (0.01-0.1) a, 
 \label{numerical factor in chaotic inf} 
\eeq
where the factor of $0.01 - 0.1$ is a numerical uncertainty. 
One example of our results is shown in Fig.~\ref{fig5}, 
where we 
set $\lambda=1$, $n=6$, $c_0 = 0.5$, $c_1 = 2$, $\abs{c_2} = 1$, and $c_3 = -1$. 
The blue curve represents the time evolution of the $B-L$ number 
after the end of inflation, 
while 
the orange dashed curve corresponds to Eq.~(\ref{numerical factor in chaotic inf}) 
with a numerical factor of $0.01$. 
The oscillation behaviour of $B-L$ number density may 
come from the effect of the oscillating inflaton through supergravity effects 
and is irrelevant for our discussion.%
\footnote{
We have investigated a possibility to generate $B-L$ asymmetry 
via this effect in Ref.~\cite{Takahashi:2015ula}. 
Note that in this paper we do not introduce a $B-L$ violating operator 
associated with the right-handed neutrino, 
so that the net $B-L$ asymmetry vanishes for this effect. 
Even if we introduce the $B-L$ violating operator, 
the resulting $B-L$ asymmetry generated from this effect is much smaller than 
that generated from ADBG. 
}
The $c_0$ dependence in our result of Eq.~(\ref{result in chaotic inf}) 
comes from the ellipticity of the dynamics of the inflaton 
in the complex plane. 
This means that $B-L$ asymmetry cannot be generated for $c_0 = 0$, 
in which case no CP odd component of the field $I$ is excited. 

One might wonder why 
there is no factor of $\abs{X}$ in our result of Eq.~(\ref{result in chaotic inf}) 
in contrast to the one in the case of hybrid inflation [see Eq.~(\ref{epsilon})]. 
Although we perform numerical calculation 
with the full supergravity potential with some kinetic interactions to derive the above results, 
we also check the same parameter dependence 
in the following toy model: 
\beq
 \ddot{\phi} + 3 H(t) \dot{\phi} + H^2 (t) \phi 
 - n a_H I X ( \phi^{*})^{n-1} &=& 0 
 \\
 \ddot{I} + 3 H(t) \dot{I} + m^2 I 
 - a_H X^* \phi^{n} &=& 0 
 \\
 \ddot{X} + 3 H(t) \dot{X} + m^2 X 
 - a_H I^* \phi^{n} &=& 0, 
\eeq
where $H(t) = 2/3 t$. 
Initial conditions are taken as 
\beq
 \phi ( t_0) = 1, 
 ~~~~
 I (t_0) = 1, 
 ~~~~
 X  (t_0) = X_0, 
 \\
 \dot{\phi} (t_0) = 0, 
 ~~~~
 \dot{I} (t_0) = i c_0, 
 ~~~~
 \dot{X}  (t_0) = 0. 
\eeq
We confirm that 
the resulting $B-L$ density 
is proportional to $c_0$ and $a_H$, 
and 
is almost independent of $X_0$.

\subsection{Baryon asymmetry}

Using the results obtained in the previous subsection, 
we calculate the baryon-to-entropy ratio such as 
\beq
 Y_b 
 &\simeq&
 \frac{2 \tilde{\epsilon} q}{23}  c_0 \frac{T_{\rm RH}}{H_{\rm osc}} 
 \lmk \frac{\phi_{\rm osc}}{\Mpl} \rmk^2 
 \label{result in chaotic 0}
 \\
 &\simeq& 
 \left\{
 \bea{ll} 
 0.005 c_0 \frac{T_{\rm RH} }{\lambda \Mpl} 
 ~~~~\text{for}~~ n = 4 
 \vspace{0.3cm} \\ 
 0.006 c_0 \frac{T_{\rm RH}}{\sqrt{\lambda H_{\rm osc} \Mpl}} 
 ~~~~\text{for}~~ n = 6, 
 \eea
 \right.
\eeq
where we assume $\tilde{\epsilon} q = 0.1$ and $\abs{c_H} = 1$ in the last line. 
For typical parameters, 
it is given by 
\beq
Y_b 
 &\simeq& 
 \left\{
 \bea{ll}
  2 \times 10^{-10} 
  \lmk \frac{c_0 T_{\rm RH}}{10^{7}  \GEV} \rmk 
  \lmk \frac{\lambda }{10^{-4}} \rmk^{-1} 
 ~~~~\text{for}~~ n = 4 
 \vspace{0.2cm}\\
 1 \times 10^{-10} 
  \lmk \frac{c_0 T_{\rm RH}}{10^{6}  \GEV} \rmk 
   \lmk \frac{\lambda}{10^{-4} } \rmk^{-1/2} 
 ~~~~\text{for}~~ n = 6, 
 \eea
 \right.
 \label{result in chaotic}
\eeq
where we use $H_{\rm osc} \simeq m_{\rm inf} \approx 10^{13} \GEV$. 
Thus, we can explain 
the observed baryon asymmetry of 
$Y_b^{\rm obs} \simeq 8.7 \times 10^{-11}$~\cite{pdg}.

Since the COBE normalisation 
of the amplitude of density perturbations 
requires that the energy scale of chaotic inflation is 
given by $H_{\rm inf} \simeq 10^{14} \GEV$ in the chaotic inflation model, 
the baryonic isocurvature constraint of Eq.~(\ref{isocurvature constraint}) 
is much severer than the case in the hybrid inflation. 
It requires that the parameter in the superpotential $\lambda$ is smaller than 
about $10^{-4}$. 
This means that the VEV of the AD field is as large as the Planck scale during inflation. 
In this case, 
the backreaction of the AD field to inflaton dynamics might be relevant. 
As a result, 
the tensor-to-scalar ratio can be consistent with the present constraint within $2 \sigma$~\cite{Yamada:2015rza}. 
Note that 
the number density of the AD field decreases with time as $\propto a^{-3}$ due to the expansion of the Universe. 
This means that its energy density decreases as $a^{-9/2}$ 
because its effective mass is of order the Hubble parameter, which decreases as $a^{-3/2}$. 
Thus its energy density never dominates that of the Universe 
and the result of Eq.~(\ref{result in chaotic 0}) is applicable even for the case of $\phi_{\rm osc} \simeq \Mpl$.

\subsection{Reheating temperature}

The inflaton can decay into the MSSM particles 
via supergravity effects. 
Its decay rate is calculated in Ref.~\cite{Nakamura:2006uc} 
and is given as 
\beq
 \Gamma_{\rm inf}^{\rm (SUGRA)} 
 = \frac{3 c_0^2}{256 \pi^3} \abs{y_t}^2 \frac{m_{\rm inf}^3}{\Mpl^2}. 
 \label{inflaton decay rate in chaotic inf} 
\eeq
This implies that 
the reheating temperature is given by 
\beq
 T_{\rm RH} \simeq 
 2 \times 10^8 \GEV 
 c_0 \abs{y_t} 
 \lmk \frac{m_{\rm inf}}{10^{13} \GEV} \rmk^{3/2}. 
\eeq
Together with Eq.~(\ref{result in chaotic}), 
we find that the observed baryon asymmetry can be explained when 
$c_0 = \mathcal{O} (0.1)$.

Note that there may be a renormalizable coupling such as 
\beq
 W \supset y X H_u H_d. 
\eeq
If $c_0$ is sufficiently small, 
the decay rate is determined by this term 
and is given by 
\beq
 T_{\rm RH} \simeq 
 6 \times 10^8 \GEV 
 \lmk \frac{ y}{10^{-6}} \rmk  
 \lmk \frac{m_{\rm inf}}{10^{13} \GEV} \rmk^{1/2}. 
\eeq
However, 
the coupling constant $y$ should be suppressed by a factor of $m_{\rm inf} / \Mpl$ 
not to affect the inflaton potential, 
so that the reheating temperature is at most $10^{9} \GEV$~\cite{Nakayama:2013txa}.

In order to kick the phase direction and generate $B-L$ asymmetry, 
we need a nonzero value of $Z_2$ breaking parameter $c_0$. 
However, 
the $Z_2$ breaking term makes the inflaton decay into gravitinos efficiently via supergravity effects 
and its decay rate is the same order with that of Eq.~(\ref{inflaton decay rate in chaotic inf}). 
Therefore, there is 
a gravitino problem from inflaton decay. 
We can avoid the problem by assuming that 
the gravitino is sufficiently heavy ($m_{3/2} \gtrsim 100 \TEV$) so as to decay 
before the BBN epoch 
and the R-parity is violated for the LSP not to overclose the Universe. 
Or, we can assume that gravitino is sufficiently light ($m_{3/2} \lesssim 2 \KEV$), 
in which case they do not overclose the Universe. 
The former possibility might be well motivated 
partly because the observed $125 \GEV$ Higgs mass favours a heavy squark mass 
of order $100 \TEV$ for a small $\tan \beta$~\cite{ArkaniHamed:2004fb, 
Wells:2004di, Hall:2011jd, Ibe:2011aa}.

\section{\label{conclusion}Discussion and onclusions}

We have investigated a new scenario 
that the Affleck-Dine mechanism works just after the end of inflation. 
The AD field stays at a large VEV by a negative Hubble-induced mass term 
during inflation 
and then starts to oscillate around the origin 
by a positive one after inflation. 
At the same time, its phase direction is kicked by an A-term 
and $B-L$ asymmetry is generated. 
Since its dynamics is determined by 
Hubble-induced terms, 
the resulting $B-L$ asymmetry is independent of parameters in low-energy SUSY models. 
This fact makes our scenario very simple. 
In particular, 
Q-balls, which sometimes form after the conventional scenario of ADBG, 
do not form in our scenario.

The A-term depends on inflation models, 
so that 
the resulting $B-L$ asymmetry does in our scenario. 
We have investigated the scenario 
and calculated the produced amount of $B-L$ asymmetry 
in F-term hybrid and chaotic inflation models in supergravity. 
We have found that 
our scenario 
requires a higher reheating temperature than the one required in the conventional scenario. 
This implies that 
ADBG works 
in larger parameter spaces than expected in the literature. 
In particular, 
in the F-term hybrid inflation model, 
the required reheating temperature 
is naturally consistent with the gravitino overproduction bounds.

The required reheating temperature is not unnaturally small 
even if 
the VEV of the AD field is so large that its backreaction to inflaton dynamics becomes relevant. 
Since the backreaction can make the spectral index and tensor-to-scalar ratio 
consistent with observations~\cite{Yamada:2015rza}, 
this is another advantage of this scenario.

\vspace{1cm}

%
\section*{Acknowledgments}
M.Y. thanks W. Buchm\"{u}ller for kind hospitality at DESY. 
This work is supported by World Premier International Research Center Initiative
(WPI Initiative), MEXT, Japan, 
the Program for the Leading Graduate Schools, MEXT, Japan, 
and the JSPS Research Fellowships for Young Scientists, No. 25.8715.
%

\vspace{1cm}



\end{document}